\RequirePackage{fix-cm}
\documentclass[twocolumn]{svjour3}

\smartqed  

\usepackage{graphicx}
\usepackage{enumitem}
\usepackage{pdflscape}
\usepackage[hidelinks]{hyperref}

\usepackage{tikz}
\usetikzlibrary{shapes,arrows}
\newcommand*\circled[1]{\tikz[baseline=(char.base)]{
		\node[shape=circle,draw,inner sep=2pt] (char) {#1};}}
	
\newcommand*\diamonded[1]{\tikz[baseline=(char.base)]{
		\node[shape=diamond,draw, minimum size=4mm,inner sep=0pt] (char) {#1};}}

\journalname{}

\begin{document}

\title{Moving Processing to Data}
\subtitle{On the Influence of Processing in Memory on Data Management}

\author{Tobias Vin\c{c}on \and Andreas Koch \and Ilia Petrov }

\institute{Tobias Vin\c{c}on \at
              Data Management Lab \\
              Reutlingen University, Germany\\
              \email{tobias.vincon@reutlingen-university.de}
           \and
           Andreas Koch \at
              Embedded Systems and Applications Group \\
              Technische Universit\"at Darmstadt, Germany\\
              \email{koch@esa.cs.tu-darmstadt.de}
           \and
           Ilia Petrov \at
              Data Management Lab \\
              Reutlingen University, Germany\\
              \email{ilia.petrov@reutlingen-university.de}
}

\date{}

\maketitle

\begin{abstract}
\textit{Near-Data Processing} refers to an architectural hardware and software paradigm, based on the co-location of storage and compute units. Ideally, it will allow to execute application-defined data- or compute-intensive operations in-situ, i.e. within (or close to) the physical data storage. Thus, \textit{Near-Data Processing} seeks to minimize expensive data movement, improving performance, scalability, and resource-efficiency.  \emph{Processing-in-Memory} is a sub-class of Near-Data processing that targets data processing directly within memory (DRAM) chips. The effective use of \textit{Near-Data Processing} mandates new architectures, algorithms, interfaces, and development toolchains. 


\end{abstract}

\section{Introduction}
Over the last decade we have been witnessing a clear trend towards the fusion of the compute-intensive and the data-intensive paradigms on architectural, system, and application levels. On the one hand, large computational tasks (e.g. simulations) tend to feed growing amounts of data into their complex computational models, on the other hand database applications execute computationally-intensive ML and analytics-style workloads on increasingly large datasets. Such modern workloads and systems are \emph{memory intensive}. And even though the DRAM-capacity per server is constantly growing, the main memory is becoming a performance bottleneck. 

Due to DRAM's and storage's inability to perform computation, modern workloads result in massive \textit{data transfers}: from the physical storage location of the data, across the memory hierarchy to the CPU. These transfers are resource intensive, block the CPU, causing unnecessary CPU waits and thus impair performance and scalability.  
The root cause for this phenomenon lies in the generally low data locality as well as in the traditional system architectures and data processing algorithms, which operate on the \textit{data-to-code} principle. It requires data and program code to be transferred to the processing elements for execution. Although \textit{data-to-code simplifies software development and system architectures}, it is inherently bounded by the \textit{von Neumann bottleneck} \cite{Kaplan:JSFI:PIM:2017,Swanson:WONDP:Micro:2014}, i.e. it is limited by the available bandwidth. Furthermore, modern workloads are not only bandwidth-intensive but also latency-bound and tend to exhibit irregular access patterns with low data locality, limiting data reuse through caching \cite{Zhang:NPP:MSPC:2013}.

These trends relate to the following recent developments:
(a)\textit{Moore's Law} is said to be slowing down for different types of semiconductor elements and \textit{Dennard scaling} \cite{Dennard1999} has come to an end. The latter postulates that performance per Watt grows at approximately the same rate as the transistor count set by Moore's Law. Besides the scalability of cache-coherence protocols, the end of \textit{Dennard scaling} is, among the frequently quoted reasons, as to why modern many-core CPUs do not have 128 cores that would otherwise be technically possible by now (see also \cite{Nikos:CS:TowardDarkSiliconInServers:2011,Muramatsu:JCDL:IfYouBuildItWillTheyCome:2004}). As a result, improvements in computational performance cannot be based on the expectation of increasing clock frequencies, and therefore mandate changes in the hardware and software architectures.
(b) Modern systems can offer much \textit{higher levels of parallelism}, yet scalability and the effective use of parallelism are limited by the programming models as well as by the amount and the types of data transfers. 
(c) \textit{Memory wall} \cite{WulfWmAandMcKee1994} and \textit{Storage Wall}. Storage (DRAM, Flash, HDD) is getting larger, cheaper but also colder, as access latencies decrease at much lower rates. Over the last 20 years DRAM capacities have grown 128x, DRAM bandwidth has increased 20x, yet DRAM latencies have only improved 1.3x \cite{Chang:PhD:2017}. This trend also contributes to slow data transfers.
(d) Modern data sets are large in volume (machine data, scientific data, text) and are growing fast \cite{Gray:SienceExponentialWorld:Nature:2006}. Hence, they do not necessarily fit in main memory (despite increasing memory capacities) and are spread across all levels of the virtual memory/storage hierarchy.
(e) Modern workloads (hybrid/HTAP or analytics-based such as OLAP or ML) tend to have low data locality and incur large scans (sometimes iterative) that result in massive data transfers. The low \emph{data locality} of such workloads leads to low CPU cache-efficiency, making the cost of data movement even higher.

\emph{In essence}, due to system architectures and processing principles, current workloads require transferring increasing volumes of large data through the virtual memory hierarchy, from the physical storage location to the processing elements, which limits performance and scalability and hurts resource- and energy-efficiency.

Nowadays, two important technological developments open an opportunity to counter these factors. 
\begin{itemize}
\item \emph{Trend 1:} Hardware manufacturers are able to \textit{fabricate combinations of storage and compute elements at reasonable costs} and package them within the same device. Consider, for instance, 3D-stacked DRAM \cite{HBM}, or modern mass storage devices \cite{Kang2013,Gu2016} with embedded CPUs or FPGAs. \emph{Cost-efficient fabrication} has been a major obstacle to past efforts.
Interestingly, this trend covers virtually all levels of the memory hierarchy: CPU and caches; memory and compute; storage and compute; accelerators -- specialized CPUs and storage; and eventually, network and compute. 
\item \emph{Trend 2:} As magnetic/mechanical storage is being replaced with semiconductor storage technologies (3D Flash, Non-Volatile Memories, 3D DRAM), another key trend emerges: \emph{the chip- and device-internal bandwidth are steadily increasing}.
(a) Due to 3D-Stacking, the chip-organization and chip-internal interconnect, the on-chip bandwidth with 3D-Stacked DRAM is steadily increasing and a function of the density. 
(b) With 3D-stacking NAND Flash chips are not only getting denser, but also exhibit higher bandwidth. For instance, \cite{Samsung1GB} presents 3D V-NAND with a capacity of 512 Gb and 1.0 GB/s bandwidth. 
Storage devices package several of those chips and connecting them over independent channels to the on-device processing element. \emph{Hence, aggregate the device-internal bandwidth, parallelism, and access latencies are significantly better than the external ones (device-to-host)}.

Consider the following numbers to gain a perspective on the above claims. The on-chip bandwidth of commercially available High-Bandwidth Memory (HBM2) is 256 GB/s per package, whereas a fast 32-bit DDR5 chip can reach 32 GB/s. Upcoming HBM3 is expected to raise that to 512 GB/s per package. Furthermore, a hypothetical 1 TB device built on top of \cite{Samsung1GB} will include at least 16 chips, yielding an aggregated on-device bandwidth of 16 GB/s, which is four times more than state-of-the-art four-lane PCIe 3.0 x4 (approx. 4 GB/s).
\end{itemize}

Consequently, processing data close to its physical storage location has become economically viable and technologically feasible over a range of technologies.

\subsection{Definitions}
\textit{Near-Data Processing (NDP)} targets the execution of data-processing operations (fully or partially) in-situ based on the above trends, i.e within the compute elements on the respective level of the storage hierarchy, (close to) where data is physically stored and transfer the results back, without moving the raw data. The underlying assumptions are: (a) the result size is much smaller than the raw data, hence, allowing less frequent and smaller-sized data transfers; or (b) the in-situ computation is faster and more efficient than on the host, thus, yielding higher performance and scalability, or better efficiency. NDP has extensive impact on:
\begin{enumerate}[nosep,nolistsep]
\item hardware architectures and interfaces, instruction sets, hardware 
\item hardware techniques that are essential for present programming models, such as: cache-coherence, addressing and address translation, shared memory.
\item software (OS, DBMS) architectures, abstractions and programming models.
\item development toolchains, compilers, hardware/software co-design.
\end{enumerate}

Nowadays, we are facing disruptive changes in the \emph{virtual memory hierarchy} (Fig. \ref{fig:neodbms:MemoryHierarchy}), with the introduction of new levels such as Non-Volatile Memories (NVM) or Flash. \textit{Yet, co-location of storage and computing units becomes possible on all levels of the memory hierarchy}.
NDP emerges as an approach to tackle the ``memory and storage walls'' by placing the suitable processing operations on the appropriate level so that execution takes place close to the physical storage location of the data. The placement of data and compute as well as DBMS optimizations for such co-placements appear as major issues. Depending on the physical data location and compute-placement several different terms are used \cite{Siegl:MEMSYS:PIMsurvey:2016,Swanson:WONDP:Micro:2014}: 
\begin{itemize}[noitemsep,nolistsep]
	\item \textit{Near-Data Processing} or \textit{In-Situ Processing} -- are general terms referring to the concept of performing data processing operations close to the physical data location, independent of the memory hierarchy level.
	\item \textit{Processing-in-Memory (PIM)}, \emph{Near-Memory Processing} -- mainly represent a paradigm where operations are executed on processing elements packaged within the memory module or directly within the DRAM chip.
	\item \emph{In-Storage Processing}, \emph{In-Storage Computing}, \emph{Process\-ing-In-Storage}  -- mostly refer to a paradigm where operations are executed on processing elements within secondary storage.
	\item \textit{Intelligent Networks}, \emph{SmartNICs}, \emph{Smart Switches} \cite{Binnig:DS:SmartNEtworks:2018} -- as the above paradigms, yet these target operation execution on processing elements within NICs or Switches. 
\end{itemize}

\emph{Challenges:} A number of challenges need to be resolved to make PIM and NDP standard techniques. Among the most frequently cited \cite{Mutlu2018,Scrbak2017,Siegl:MEMSYS:PIMsurvey:2016} are: the host processor and the PIM processing elements (as well as their instruction sets and architectures); the memory hierarchy, the memory model, established techniques such as TLB and address translation, cache-coherence, synchronization mechanisms and shared state/memory techniques; interconnect, communication channels, interfaces and transfer techniques; programming models and abstractions; OS/DBMS integration and support.

\subsection{Historical Background}
The concept of \emph{Near-Data Processing} or \emph{in-situ processing} is not new. Historically it is deeply rooted in the concept of \textit{database machines} \cite{DeWitt:CACM:DatabaseMachines:1992,Boral:DatabaseMachines:1989} developed in the 1970 and 1980s. \cite{Boral:DatabaseMachines:1989} discuss approaches such as processor-per-track or processor-per-head as an early attempt to combine magneto-mechanical storage and simple computing elements to process data directly on mass storage and reduce data transfers.  Besides reliance on proprietary and costly hardware, the I/O bandwidth and parallelism are claimed to be the limiting factor to justify parallel DBMS \cite{Boral:DatabaseMachines:1989}. While this conclusion is not surprising, given the characteristics of magnetic/mechanical storage combined with Amdahl's balanced systems law \cite{Gray:RulesOfThumb:ICDE:2000}, it is revised with modern technologies. Modern semi-conductor storage technologies (NVM, Flash) are offering high raw bandwidth and high levels of parallelism. \cite{Boral:DatabaseMachines:1989} also raises the issue of temporal locality in database applications, which has already been questioned earlier and is considered to be low in modern workloads, causing unnecessary data transfers. Near-Data Processing presents an opportunity to address it.

The concept of \textit{Active Disk} emerged towards the end of the 1990s. It is most prominently represented by systems such as: Active Disk \cite{Acharya:ASPLOS:ActiveDisc:1998}, IDISK \cite{Keeton:SigmodRec:IDISK:1998}, and Active storage/disk \cite{Riedel:VLDB:ActiveStorage:1998}. While database machines attempted to execute fixed primitive access operations, \textit{Active Disk} targets executing application-specific code on the drive. Active storage \cite{Riedel:VLDB:ActiveStorage:1998} relies on processor-per-disk architecture. It yields significant performance benefits for I/O bound scans in terms of bandwidth, parallelism and reduction of data transfers. IDISK \cite{Keeton:SigmodRec:IDISK:1998}, assumed a higher complexity of data processing operations compared to \cite{Riedel:VLDB:ActiveStorage:1998} and targeted mainly analytical workloads and business intelligence and DSS systems. Active Disc \cite{Acharya:ASPLOS:ActiveDisc:1998} targets an architecture based on on-device processors and pushdown of custom data-processing operations. \cite{Acharya:ASPLOS:ActiveDisc:1998} focuses on programming models and explores a streaming-based programming model, expressing data intensive operations, as so called \textit{disklets}, which are pushed down and executed on the disk processor.

\begin{figure}
  \includegraphics[scale=0.63]{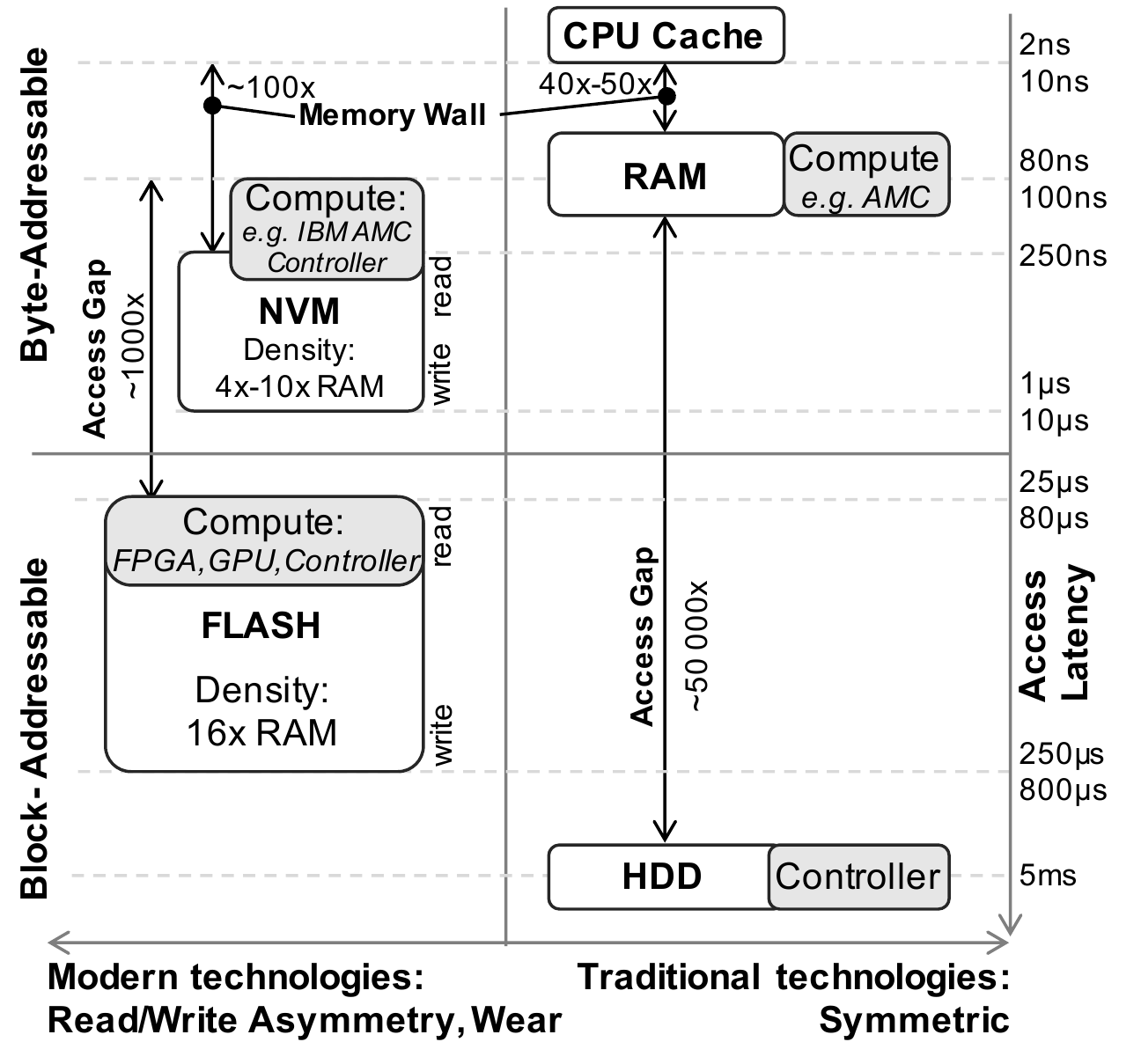}
  \caption{Complex Memory Hierarchy. Co-location of storage and compute.}
  \vspace{-5pt}
  \label{fig:neodbms:MemoryHierarchy} 	
\end{figure}

As a result of recent developments and \emph{Trend 1}, the memory hierarchy nowadays is getting richer and incorporates new levels. Also, processing elements of different types are being co-located on each level (Fig. \ref{fig:neodbms:MemoryHierarchy}). Hence, NDP has diversified depending on the level. \cite{Agerwala:DataIntensive:ICPP:2014} acknowledges that computing should be done on the appropriate level of the memory hierarchy (Fig. \ref{fig:neodbms:MemoryHierarchy}) and that, in the general case, it will be \emph{distributed} along all levels and is \emph{heterogeneous} because of the different types of processing elements involved.

Research combining memory technologies with the above ideas, often referred to as Processing In-Memory (PIM), is very versatile, and likewise not a new idea. In the late 1990, \cite{Patterson:IRAM:Micro:1997} proposed IRAM as a first attempt to address the memory wall, by unifying processing logic and DRAM. \cite{Kaxiras:MLD:VectorIRAM:1997} proposed moving computation to data rather than vice versa to reduce data movement. This idea gave rise to the concept of \emph{memory-centric computing} \cite{Minutoli:RadixSortEmu1:WONDP:2015} or \emph{data-centric computing} \cite{Agerwala:DataIntensive:ICPP:2014} and found also application in various computer science technologies besides data management systems \cite{Boroumand2018}. \cite{Balasubramonian:MSSC:MakingTheCaseForFeatureRichMemorySystems:2016} provides an excellent overview of modern PIM techniques.
With the advent of 3D-Memories, PIM is said to become commercially viable \cite{Torrellas:FlexRAMretro:ICCD:2012} (see Section \ref{sec:technologies} for more details). The possible PIM performance improvement is illustrated in \cite{Zhang:NPP:MSPC:2013}, where 50\% latency (and 77\% execution time) improvements are reported under a latency sensitive workload; and 4x bandwidth improvement under a bandwidth-sensitive workload in PIM settings.

Nowadays, the  NDP builds upon ActiveDisk/Storage ideas in terms of \emph{processing-in-storage} gain significant attention in terms of intelligent storage concepts such as: \emph{SmartSSDs}, \emph{In-Storage Processing/Computing}. With growing datasets that do not fit in memory, many data- and compute-intensive (e.g. selections, aggregations, joins or linear algebra) operations can be performed directly within mass storage as a result of \emph{Trend 2}. In terms of NDP performance, \cite{Moon:InfSci:ISC:2016} reports 7x and 5x improvement for scans and joins and energy savings of up to 45x.

\emph{Accelerator-based computing.}
Based on the observation that the difficulties of constructing general hardware can be avoided by constructing dedicated cards with new designs and connecting them to the cost over standard interfaces gave rise to the so-called \emph{accelerator based computing}. A good overview of the emerging DBMS research, which examines using a GPU as co-processor, is provided in \cite{Bress:GPUcoproc:IS:2013}. \cite{Teubner:IBEX:VLDB:2014} is an excellent example of NDP on FPGA-based accelerators demonstrating performance improvement of 7x to 11x under TPC-H workloads.

\subsection{PIM Problem Space}
\begin{figure*}
	\includegraphics[clip,trim=0.3cm 0cm 0cm 10cm,scale=0.83]{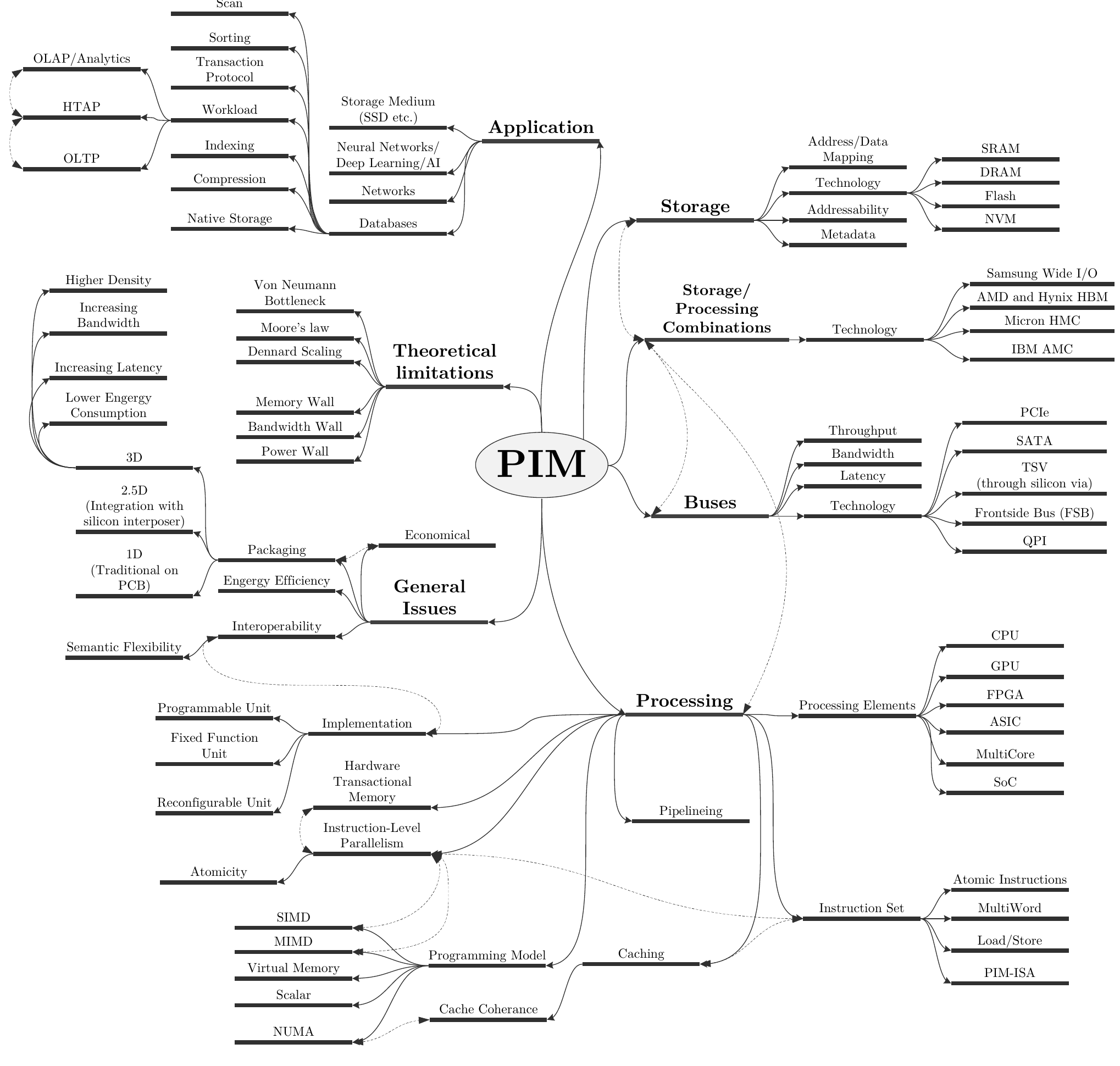}
	\caption{Problem space of Processing In-Memory.}
	\vspace{-5pt}
	\label{fig:problemspace} 	
\end{figure*}

NDP and PIM impact the foundation of established computing and architectural principles. Naturally, the problem space of such paradigms involves a wide range of aspects. Those covered in the present survey are depicted in Figure \ref{fig:problemspace}. We consider present limitations as well as technological and fabrication trends that lead us to believe that current PIM-efforts represent a breakthrough given the current state-of-the-art. Modern workloads, systems as well as developments, and data processing, and analytics are major factors in favor of PIM. Last but not least, aspects such as \emph{interconnects, processing elements, instruction sets, memory, computing, and synchronization techniques} as well as the \emph{programming models} play a central role in the current survey.

\section{Technological Advances}
\label{sec:technologies}

\subsection{Storage}

The research about storage technologies has been dramatically involved in the semiconductor industry during the last decades. Mainly two types are of interest -- Flash and lately Non-Volatile Memories.

\subsubsection{Flash}
With the advent of the Electrically Erasable Programmable Read-Only Memory (EEPROM) the first NOR flash cell by Toshiba \cite{Masuoka1985} was presented in the mid 1980s. This gave rise to a new type of purely electrical storage technology (without any mechanical moving parts) with read/write latencies an order of magnitude lower than traditional magnetic drives (see Table \ref{tab:comparisonstorage}). Almost two years later, Toshiba's engineer Masuoka introduced the Flash EEPROM as a NAND structure cell \cite{Masuoka1987}, enabling to produce smaller cell sizes without scaling the device dimensions.

In the subsequent years, the trend towards structure size reduction\footnote{Smaller structures refer to shrinking dies, mainly due to shrinking transistors sizes. The process is repeatedly defined by ITRS \cite{ITRS2011}, e.g. 2018: 7nm, 2017: 10nm, 2014: 14nm, 2012: 22nm, 2010: 32nm, 2008: 45 nm, ...} and advanced fabrication processes drove the competition among the flash chip vendors. As a result, the current floating gate transistors, as one essential component for flash cells, can differentiate between multiple states of electrical charge to increase the data density. While Single-Layer Cells (SLC) are only able to store one single bit per cell, Multi-Layer Cells (MLC) \cite{Park2015} or Triple-Layer Cells (TLC) \cite{Kang2017} can persist two or three bits per cell. Recently, even Quad-Layer Cells (QLC) \cite{Liu2018} are introduced. Since the miniaturization of structures represents a major obstacle, as physical limits are reached, \emph{stacking} approaches were recently applied. As a result, stacked planar flash chip topologies (2D-NAND) were recently replaced by the so-called 3D-NAND \cite{Silvagni2017}. This could be done either horizontally \cite{Sakuma2013} or vertically \cite{Parat2015,Park2015,Kang2017} to lower the production costs, increase capacities, and reduce the aggregated SSD power consumption.

\begin{table*}
	\centering
	\caption[Comparison of storage technologies]{Comparison of storage technologies~\cite{ITRS2011}}
	\label{tab:comparisonstorage} 
	\begin{tabular}{p{3.4cm}p{1.1cm}p{1.1cm}p{1.6cm}p{1.4cm}p{1.9cm}p{1.1cm}}
		\hline\noalign{\smallskip}
		& DRAM & PCM & STT-RAM & Memristor & NAND Flash & HDD \\
		\noalign{\smallskip}\hline\noalign{\smallskip}
		Write Energy [pJ/bit] & 0.004  & 6 & 2.5 & 4 & 0.00002 & $10x10^9$\\
		\noalign{\smallskip}
		Endurance 			  & $>10^{16}$ & $>10^{8}$ & $>10^{15}$ & $>10^{12}$ & $>10^{4}$ & $>10^{4}$\\
		\noalign{\smallskip}
		Page size 			  & 64B  & 64B & 64B & 64B & 4-16KB & 512B\\
		\noalign{\smallskip}
		Page read latency 	  & 10ns & 50ns & 35ns & $<$10ns & $\sim$25us & $\sim$5ms \\
		Page write latency 	  & 10ns & 500ns & 100ns & 20-30ns & $\sim$200us & $\sim$5ms \\
		Erase latency 		  & N/A  & N/A  & N/A  & N/A  & $\sim$2ms  & N/A \\
		\noalign{\smallskip}
		Cell area [F$^2$] 	  & 6 & 4-16 & 20 & 4 & 1-4 & N/A\\
		\noalign{\smallskip}\hline
	\end{tabular}
\end{table*}

\subsubsection{Non-Volatile Memory}
In parallel, the research on novel non-volatile memory technologies like Spin-Transfer Torque Random Access Memory (STT RAM) \cite{Huai2008}, Phase-Change Memory (PCM) \cite{Lee2009}, Magnetoresistive Random Access Memory (MRAM) \cite{Wong2012} or Resistive Random Access Memory (RRAM) began. Concrete technologies and devices were recently announced by semiconductor vendors: Intel and Micron's PCM \cite{Villa2010} and 3D XPoint \cite{3DXPoint}, Samsung's PCM \cite{Choi2012}, HPE's Memristor \cite{Strukov2008} or Toshiba and Sandisk's RRAM \cite{Scheuerlein2013}. They are subsumed under the term \emph{Non-Volatile Memories (NVM)} or \emph{Storage Class Memory (SCM)}. NVM characteristics differ from conventional storage technologies like Flash or DRAM: like DRAM they are byte-addressable, yet the read/write latencies are 10x/100x higher than DRAM  (see Table \ref{tab:comparisonstorage}). Unlike DRAM, NVM operations are \emph{asymmetric}, i.e. reads are much faster than writes. Like flash, cells wear out with the number of program cycles, making it necessary to employ wear-leveling approaches (like a Flash-Translation Layer (FTL)) to distribute writes evenly across all cells and ensure even wear over time. 

\subsection{Processing Elements}
The invention of mass-produced processing units dates back into the late 1960s with the foundation of vendors like Intel or AMD. In 1971, the first microprocessor 4004 was announced by Intel \cite{Faggin1996}, comprising 16 Read-Only Memory (ROM) and 16 Random-Access Memory (RAM) chips. Since then, the processing units evolved dramatically. 

Nowadays Intel Skylake-SP Central Processing Units (CPU) comprise of up to 28 cores, clocked with 3.6 GHz, and can address 1.5 terabytes of memory. AMD's counterpart, the Zen-based Epyc processor has even up to 32 cores per CPU. Besides the classical ALUs, current CPUs include also multiple caches, buffers, and vector units. A conventional server can be equipped with multiple of such CPUs (typically 4 to 8), resulting in an extremely high parallelism. 

Having even more cores per processing unit, Graphical Processing Units (GPU) became a common accelerator for various algorithms of many applications besides graphical processing calculations. Especially tasks like matrix computation, used in artificial intelligence or robotics, are perfectly fitted to the vectorized SIMT fashion of a GPU. Lately, Application-Specific Integrated Circuits (ASIC) like the Google's TensorFlow Processing Units (TPU) have become more and more prominent as their performance for specific workloads is immense. 

However, ASICs can perform only algorithms defined during the development and cannot be changed afterwards or during runtime. For this reason, more flexible Field-Programmable Gate Arrays (FPGA) or Coarse Grained Reconfigurable Architectures (CGRA) are applied in such diverse workloads, but still have an extreme parallelism because of their ability to maintain multiple dynamic and elastic pipelines.

\subsection{Packaging and 3D Integration}

\begin{figure}
  \includegraphics[scale=0.55]{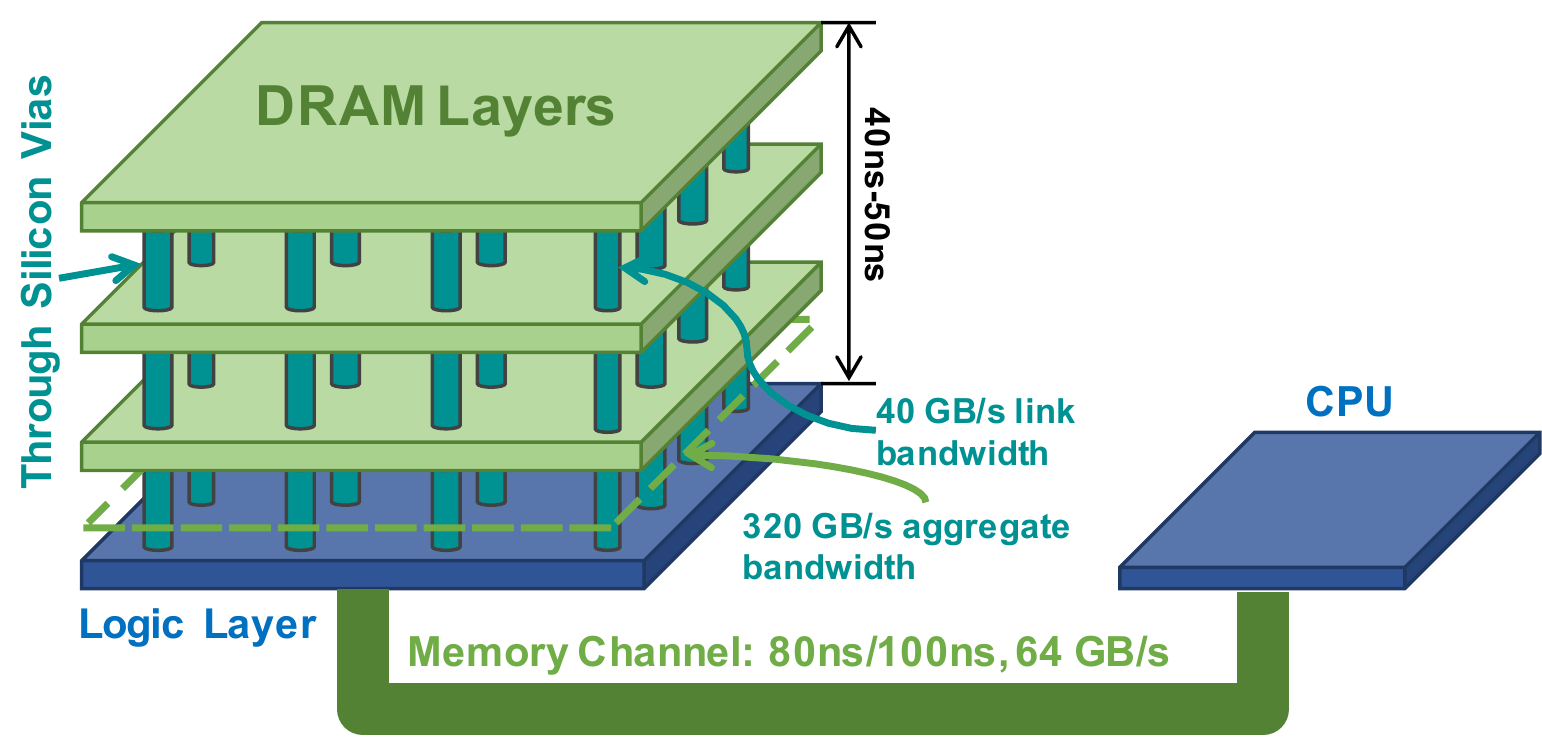}
  \caption{Architecture of a 3D-stacked DRAM (based on \cite{Mutlu2018,Scrbak2017}).}
  \vspace{-5pt}
  \label{fig:3Dmem} 	
\end{figure}

With the advent of fabrication processes in the semiconductor industry the ability to manufacture highly integrated circuits allowed for shorter wire paths by placing heterogeneous elements on the same silicon die. Such systems, partially referred as Very Large Scale Integration (VLSI), revolutionized the computer science, are basis for all modern technology, and was awarded with the Nobel prize to its inventor Jack S. Kilby in 2000 \cite{ICNobelprize}. The packaging arrangements varied over the years from Multi-chip Modules (MCM) over System-on-Chip (SoC) to System-in-Package (SiP) or even Package-in-Package (PiP). All those have in common that the die or dies are mounted in the package in a single plane and therefore are known as 2D devices.

However, with the growing number of transistors per chip (see Moore's law \cite{Schaller1997}) and the end of faster and more efficient transistors (see end of Dennard scaling \cite{Dennard1999}) led to micro-architectures with a larger number of separate processing elements (e.g. cores or even processing elements of different types like GPU or FPGA), instead of a substantial increase of single-core performance. Along the lines of Moore's law \cite{Schaller1997}, transistors would need to shrink to a scale of a handful of atoms by 2024 \cite{Shalf2015}, under the traditional 2D silicon lithographic fabrication process. Separate processing elements though, have much higher I/O requirements (e.g. memory bandwidth) than single cores. These I/O requirements have proven to be difficult to fulfill even in chip packages with thousands of pins. However, they can be achieved by stacking chips either directly on top of each other (3D) or on top of a passive interposer die (2.5D) and perform connections not through pins/balls but using much smaller, but far more numerous Through-Silicon Vias -- TSVs (Figure \ref{fig:3Dmem}). Using 10 000 of TSVs, I/O bandwidths in the order of terabytes per second can be achieved. Furthermore, through shorter wires, like TSVs, the system-wide power consumption can be reduced, decreasing the impact of heat development or parasitic capacitance. However, with growing complexity of 3D integrated circuits same effects re-emerge as challenges \cite{Kim2009}. 

3D integrated circuits enable fabrication of heterogeneous systems, including logical processing and persistence, on the same chip \cite{Borkar2010}. Well known examples for such heterogeneous systems are the High Bandwidth Memory (HBM) from AMD and Hynix \cite{Lee2014}, Samsung's Wide I/O \cite{Kim2012}, the Bandwidth Engine (BE2) \cite{Miller2011}, the Hybrid Memory Cube (HMC) developed by the Micron and Intel \cite{Pawlowski2011} or IBM's Active Memory Cube (AMC) \cite{Nair2015}.

\subsection{Interconnect and Buses}
Modern computer architectures include a large number of different bus systems. Firstly, there are peripheral bus systems to connect the host bus adapter to mass storage devices (e.g. HDDs or SSDs). Standards like SCSI, FibreChannel and SATA are omnipresent but suffer from the poor bandwidth performance improvements. For instance, the first version of SATA in 2003 is able to transfer only 1.5 Gbit/s, yet the latest version from 2008 has only improved by a factor of four. Secondly, there are expansion buses, which connect various devices to the host system (e.g. graphic cards, accelerators, or storage devices). Standards like PCIe significantly increased their bandwidth from 4 GB/s in its first version to about 32 GB/s using 16 lanes in the recent PCIe v4.x.

The interconnects between the CPUs and the memory, should also be considered besides these above bus types. During the 1990s and 2000s the front-side bus (FSB) of Intel and AMD connected the CPU with the northbridge in the computer architectures. With the low throughput of around 4-12 GB/s they got replaced by the Quick Path Interconnect (QPI) or the HyperTransport (HT) interfaces in modern systems. QPI operates on 3.2 GHz and has a theoretical aggregated throughput of 25.6 GB/s. HT doubles this, because it directly uses 32 instead of 16 data bits per link. AMD's newest on-chip interconnect architecture, Infinity Fabric (IF), is even specified to transfer about 30 to 512 GB/s. 

As a consequence, there is a \emph{significant bandwidth gap} between the on-chip bandwidth and the off-chip bandwidth (i.e DRAM-to-CPU) as depicted in Figure \ref{fig:3Dmem}. Furthermore, due to the RAS/CAS interface and the internal DIMM module organization DRAM offers limited parallelism, while increasing the number of DIMMs per channel typically decreases performance \cite{Swanson:WONDP:Micro:2014}.

\subsection{Summary}
The advance in computer technologies over the last decades is remarkable. Especially the storage and memory chips have increased their volumes per area dramatically. CPUs as processing elements have reached their limits in clock frequencies but just started to scale horizontally over multiple cores, resulting in an immense parallelism. Additionally, new processing technologies become more and more mainstream to implement in nowadays data centers and are perfectly fitted for modern problems like AI.

With the exception of 3D integration, buses between storage/memory and processing elements have slightly evolved in comparison to the remainder. Newer bus systems have promising throughputs but cannot withstand the foreseeable workloads. As a consequence, PIM is a promising alternative to scaling the bus bandwidth and achieving low latencies needed by modern workloads and applications.

\section{Impact on Computer Architecture, OS, and Applications}
\label{sec:computerarchitecture}

Despite all research and technological advancements, especially those regarding physical boundaries of fabrication processes (e.g. more transistors per area) or run time properties (e.g. heat dissipation or power consumption), the switch from the \textit{data-to-code} to the \textit{code-to-data} impacts the computer and systems architecture (operating system or DBMS) as well as the applications running on top of them. Concepts like virtual memory and address translation have to be adapted to novel computational and programming models. To this end, instruction sets of processing and storage elements need reconsideration, coherency protocols need to be revisited, and the workload is distributed across the entire system. Ideally, cross layer optimizations would touch multiple layers of the system hierarchy and thus, mitigate performance bottlenecks of traditional abstractions or concepts while focusing likewise on latency and interoperability \cite{Swanson}.

The following section classifies PIM research from the last three decades, with respect to: workload distribution/partitioning, instruction sets, computational/ programming model, addressing, buffer/cache management, and coherency. An overview of the evaluated approaches and their classification is given in Table \ref{tab:stateofresearch}. Yet, there are several more approaches present in research \cite{Zhang2018,Seshadri2017,Nai2017,Tsai2018,Tang2017,Gao2015,Xie2017}.

\subsection{Computational and Programming Model}
\label{sec:computationalmodel}
\begin{figure}
  \includegraphics[scale=0.42]{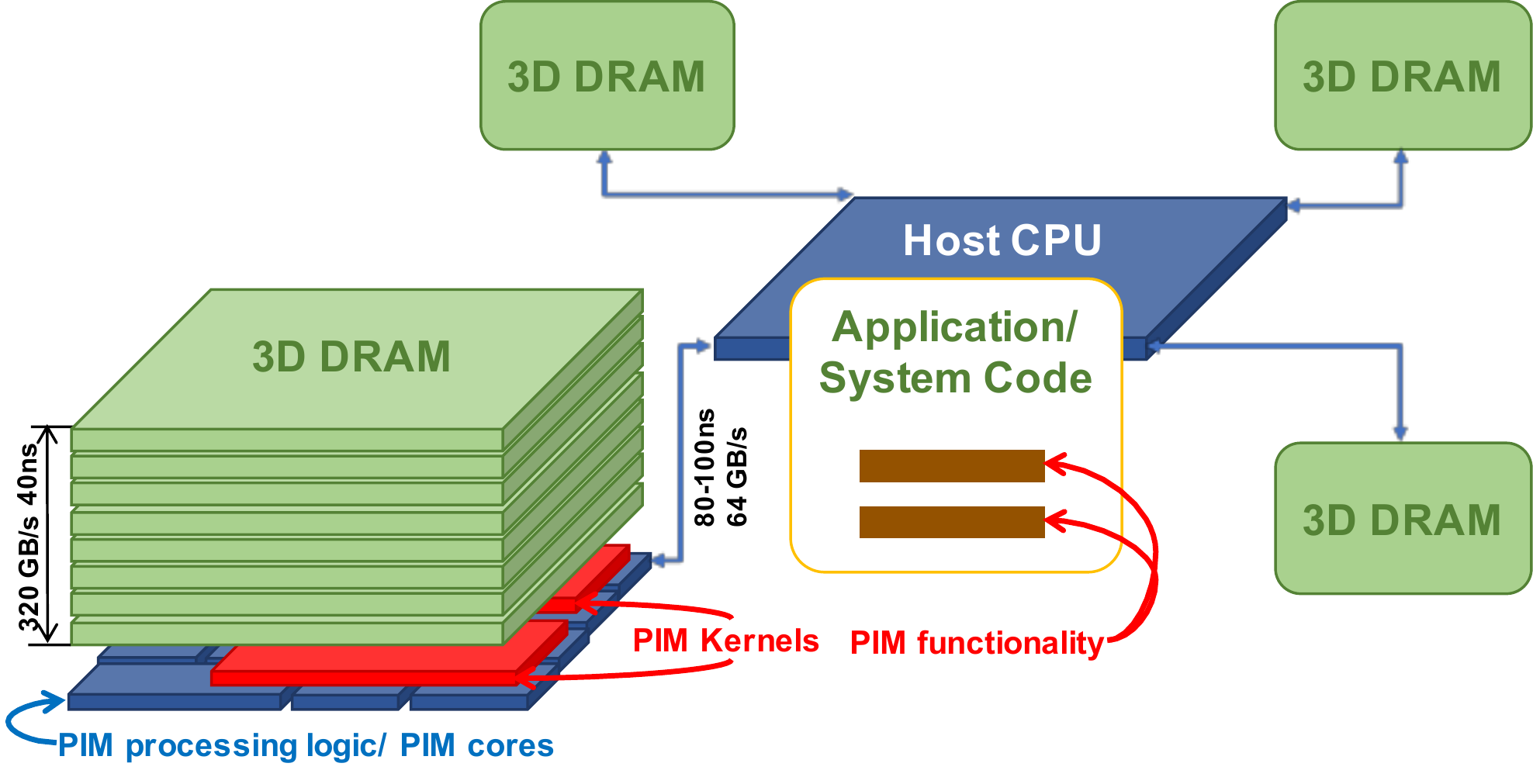}
  \caption{PIM Architecture with 3D-stacked DRAM.}
  \vspace{-5pt}
  \label{fig:PIMmodel} 	
\end{figure}

PIM architectures place \emph{PIM processing logic} on the \emph{logic layer} within the DRAM chip itself (Figure \ref{fig:3Dmem}) or on a processing element within the memory module (Figure \ref{fig:PIMmodel}). The offloaded PIM processing logic is typically referred to as \emph{PIM cores} or \emph{PIM engines} \cite{Mutlu2018,Scrbak2017}. Currently, PIM cores have limited use, yet many research proposals, making efficient use of it, have appeared recently \cite{Mutlu2018}. Depending on the architecture, these ``range from fixed-function accelerators to simple in-order cores, and to reconfigurable logic'' \cite{Mutlu2018}. A broad taxonomy of the different PIM Cores functionality is presented in \cite{Loh2013}. 

The PIM cores execute only when application/code is spawned by the CPU on the PIM processing logic. The offloaded parts of the system/application on the PIM core are typically referred to as \emph{PIM kernels} (Figure \ref{fig:3Dmem}). PIM kernels vary significantly in their scope and functionality. Many recent research of PIM architectures follow similar models for CPU-PIM (core and kernel) interactions in terms of interfaces, techniques, and programming models. 

\subsubsection{PIM Computational Models}

Processing data closer to storage or even memory allows for concurrent processing of higher data volumes. Therefore computational or programming models like vector processing and data parallelism based on Single Instruction Multiple Data (SIMD) or even Multiple Instruction Multiple Data (MIMD) play an important role. Already in 1994, P. Kogge presented the \emph{EXECUBE} \cite{Kogge1994} for massively parallel programming. \emph{EXECUBE} is fabricated with processing logic and memory side-by-side on a single circuit. It comprises 4 MB DRAM, which is equally partitioned in a logic array implementing 8 complete 16-bit CPUs. These 8 processing elements can obtain their instructions from their memory subsystems and run in a MIMD mode or can be addressed from the outside, utilizing a SIMD Broadcast Bus, by sending instructions directly into the CPU's instruction register.

Another approach is presented by \cite{Gokhale1995}, who divided the MIMD and SIMD processing on different parts of their \emph{Terasys} system. Given their new programming language, \emph{data-parallel bit C}, conventional instructions are executed by the SPARC processor while data parallel operands are promoted to the ALU within the memory. This allows executing applications, which are well suited for SIMD processing without penalizing conventional application logic. Computational RAM (CRAM), Elliott et al. \cite{Elliott1999,Elliott2008} follow an approach similar to \cite{Gokhale1995}, which can function either as a conventional memory chip or as a SIMD processing unit. Benchmarks, comparing the CRAM with a setup based on the SPARC processor, show an impressive speed-up of up to 41 times.

A more flexible approach, called \textit{Active Pages} \cite{Oskin1998}, has been presented by a research team of the UC Davis. \textit{Active Pages} \cite{Oskin1998} introduce a novel computational model, where each page consists of data and a set of associated functions. Those functions can be bound during runtime to a group of pages and be applied on data located within these pages. The implementation is based on Reconfigurable RAM, combining DRAM with reconfigurable logic. 

As such combinations of general purpose processors or vector accelerators with memory chips are non-trivial to fabricate, ProRAM \cite{Wang2015} was proposed to leverage existing resources of NVM devices to implement a lightweight in-memory SIMD-like processing unit. ProRAM is based on the Samsung's PCM architecture \cite{Lee2009} to reuse and instrument the Data Comparison Write unit in combination with further surrounding peripheral units for processing, e.g. additions, subtractions or scans.

The near-DRAM accelerator (NDA) \cite{Farmahini-Farahani2015} introduces a completely new programming model, utilizing modern 3D stacking. Similar to \cite{Riedel1999}, the application code is profiled and analyzed for data-intensive kernels with long execution times, which is then converted into hardware data-flows. These can be executed by CGRA units (Coarse-Grained Reconfigurable Architecture) located near-DRAM in a highly parallel fashion. 

The architecture of the Active Memory Cube (AMC) \cite{Nair2015} takes this concept one step further as it implements a balanced mix of multiple forms of parallelism such as multithreading, instruction-level parallelism, vector and SIMD operations. To facilitate the programming model a special AMC compiler is necessary to generate AMC bitcode out of the user-identified code sections and data regions. Several compiler optimization techniques (e.g. loop blocking, loop versioning, or loop unrolling) are utilized to exploit all forms of parallelism. Same concept can also be found in other approaches like \textit{TESSERACT} \cite{Ahn2015} or the Mondrian Data Engine \cite{Drumond2017}.

\subsubsection{PIM Programming Models}

One of the challenges towards a wide-spread use of PIM lies in the appropriate \emph{programming models}. Many system designs treat the PIM processing logic (PIM core) as a \emph{co-processor}. Hence, many PIM programming models are rooted in accelerator-based computing approaches. \cite{Scrbak2017,Siegl:MEMSYS:PIMsurvey:2016} consider \emph{MapReduce} as a suitable model in compute-intensive environments such as HPC. Furthermore, frameworks/models such as \emph{OpenMP} and \emph{OpenACC} are considered good candidates. They, however, need specific PIM extensions to cover broader classes of PIM operations. Moreover, \emph{OpenCL} is considered a viable programming model alternative \cite{Zhang:NPP:MSPC:2013}. It is based on the heterogeneous computing characteristics of PIM and the typically data-parallel operations. Interestingly, programming models for data processing and database functionality for PIM (similar to the ones designed for GPGPU accelerators \cite{Bress:GPUcoproc:IS:2013}) are still an open topic.

\cite{Mutlu2018,Boroumand:CAL:LazyPIM:2017} argues that existing programming models should be preserved for PIM to simplify application development and allow for easy spread of PIM architectures. \cite{Mutlu2018,Boroumand:CAL:LazyPIM:2017,Zhang:NPP:MSPC:2013,Scrbak2017} claim that elementary techniques such as a single virtual memory space (and address translation) as well as cache coherence should be preserved for PIM. Furthermore, many authors \cite{Mutlu2018,Zhang:NPP:MSPC:2013,Scrbak2017} seem to agree that PIM inherently should rely on \emph{non-uniform memory access} techniques and system architectures. 

PIM infrastructures and development toolchains are major aspects towards broader PIM proliferation. PIM infrastructures are intrinsically of heterogeneous compute nature: \cite{Scrbak2017} for instance assumes ARM-like PIM cores, whereas \cite{Zhang2014} focuses on GPU-based PIM-Cores, while HMC-based alternatives rely on FPGAs. Domain-Specific Languages (DSL) and highly optimized libraries as well as compiler infrastructures have proven to allow efficient development over the set of the above technologies and hardware/software co-design.
Furthermore, suitable debugging, monitoring, and profiling tools are essential for PIM-enabled architectures, yet they are still referred to as future work.

For In-Storage Processing, where the processing unit is near the storage chips on a PCB, the computational model is usually hidden behind a well-defined host-to-device interface. \cite{De2013,Cho2013,Minutoli2015,Gu2016} are only a few examples of such systems. Their interface is designed either for general-purpose applications or specific workloads but offers the opportunity to address the main advantage of both, PIM and NDP, the reduction of memory transfers.

\subsection{Processing Primitives and Instruction Sets}
\label{sec:instructionset}
Conventional systems are mainly based on a \emph{load and store} semantic, where cachelines are transferred from the memory into the caches or registers of the processing unit and vice versa. The low-level instructions are executed on the cacheline data and the result is evicted to the memory subsystem (store), when advised. Thereby, the available instruction set is determined by the processing unit's architecture, often referred as Instruction Set Architecture (ISA). Today one can divide ISAs in Reduced Instruction Set Computer (RISC) and Complex Instruction Set Computer (CISC). These are either standardized by foundations like the RISC-V \cite{RISCV} and/or extended by vendors, e.g. Intel introduced the Intel Instruction Set Architecture Extensions (Intel AVX) \cite{IntelAVX} for vector-based programming models like SIMD. 

With PIM in place, the interface to the memory needs to be extended. Depending on the depth of PIM integration the architecture and the type of interconnect (host-to-device/host-to-memory) the interface may take different forms. Either single primitives, implemented directly in the PIM processing logic and controlled by traditional higher-order instructions are evaluated, or the existing instruction sets are extended to have novel ways of managing the processing capabilities in the memory.

Terasys \cite{Gokhale1995} -- a pioneering PIM system -- reused patterns from the existing CM2 Paris instruction set \cite{ThinkingMachines}. Seshadri et al. \cite{Seshadri2015} recommend modifing applications for their bulk bitwise AND and OR DRAM by making use of specific instructions. Modifications are limited to a preprocessor and to specific libraries, which exploit the available hardware acceleration and are shared by many applications. 

Often a compiler is necessary to cover the complexity of those low-level instructions for software developers. For this purpose, \cite{Hall1999,Draper2002} focus on the Stanford SUIF compiler system to hide the complexity of DIVA, which has high similarities to a distributed-shared-memory multiprocessor. It supports their special \textit{At-the-Sense-Amps Processor} instructions that are seamlessly integrated into the PIM backend. Likewise, \cite{Hadidi2017} analyze within their proposed CAIRO the advantages of compiler-assisted instruction-level PIM offloading and thereby examine an double in performance for a set of PIM-beneficial workloads.

Higher level of parallelism can also be achieved by, e.g. long-instruction words (LIW) as proposed by AMC \cite{Nair2015}. The ISA intends to have a vector length, which is applied on all vector instructions with the LIW. It can be applied either directly within the instruction or by obtaining it from a special register. By this, AMC can express three different levels of parallelism: parallelism across functional pipelines, parallelism due to spatial SIMD in sub-word instructions, and parallelism due to temporal SIMD over vectors \cite{Nair2015}.

Often, PIM instructions are also hidden behind an interface. For instance, JAFAR defines its select instruction as an API call with start address of the virtual memory address and further parameters such as range\_low and range\_high as inclusive bounds for range filters \cite{Xi2015,Babarinsa2015}. Similar considerations apply to many proposed in-storage or accelerator-based approaches, which focus on the instructions from an application point of view but not the seamless integration into the computer architecture itself \cite{Kang2013,Koo2017,Ist2017,Seshadri2014}. For example, \cite{Wu2014} introduces a flexible and extensive database specific instruction set for an accelerator, but reaches the out-of-chip bandwidth limits, which is a clear evidence for the necessity of PIM.

One of the most advanced processing interface is the Expressive Memory (XMem) introduced in \cite{Vijaykumar2018}. Besides many other optimizations like cache management and compression, placement, and prefetching, the authors propose a new hardware-software abstraction called \emph{Atom}. An Atom consists of three key components, \emph{Attributes} for higher-level data semantics, \emph{Mapping} to describe the virtual address range, and \emph{State} indicating whether the Atom is currently active or inactive. These can be manipulated by issuing specific calls. Atoms are planned to be interpretable by all layers of the computer architecture and therefore constitute a cross-layer interface. Thereby, it is possible to interact with Atoms on every level and execute PIM operations in hardware or in software.

\subsection{Memory Management and Address Translation}
\label{sec:memorymanagement}
The question of addressing and address management is directly related to the ISA and the computational model. This includes the distribution of the usable storage or memory into address spaces and the way of addressing it, i.e. directly executing on physical address or by exploiting virtual addresses to the host system and managing any kind of page table. In-storage solutions \cite{Fitch2009} or large-scale platforms \cite{Gu2016}, which evaluate PIM-like problems, often rely on traditional storage APIs that use immutable logical address and an address granularity of whole blocks, which is not sufficient for memory. 

To this end, a few PIM approaches set their focus differently, but take advantage of the existing address management of modern CPUs. For instance, JAFAR's API \cite{Xi2015,Babarinsa2015} must be called for every page since the address translation service is managed by the host. A similar approach is pursued by \cite{Ahn2015ISCA}, which supports virtual memory as their PIM-enabled instructions are part of a conventional ISA. Hence, they avoid the overhead of adding address translation capabilities into the memory and leverage existing mechanisms for handling page faults. A slightly different approach is to utilize memory mappings of address ranges within the executing program. Commands, notifications, and results can be processed by writing and reading predefined addresses \cite{Gokhale2015}. The AMC \cite{Nair2015} solves the same problem by dividing the classical load/store subsystem, which is responsible to perform read and write access to the memory, and the computational subsystem that performs transformations of data.

Another approach is to implement a page table within the PIM modules instead of reusing the host's one. FlexRAM \cite{YiKang1999} is likewise based on virtual memory, which shares a range with the host system, but nevertheless, the programmer can explicitly specify how the data structure is distributed on the different physical locations. The memory module takes care of the virtual to physical address translation, which in the case of FlexRAM organized with base and limit page number for each data structure. These contiguous memory allocations minimize the wasted memory space and reduce the time necessary for sequential mapping traversal. Furthermore, to reduce TLB invalidations by page replacements, shared pages must be pinned in the beginning of each program to ensure that only private pages can be replaced. DIVA \cite{Hall1999,Draper2002} extends the approach of a fixed relationship between virtual and physical address since it was determined to be too restrictive. In \cite{YiKang1999}, each PIM core contains a translation hardware, but the tables are managed by the host to facilitate that any virtual page can reside on any PIM. Because the PIM core needs to rapidly determine if an address is local to its own memory bank, PIM cores additionally maintain translations for those virtual pages currently residing on it. Non-local pages can be addressed by querying the global table residing on the host system. There are also completely new concepts for page tables, such as the region-based page table of IMPICA \cite{Hsieh2016} to leverage the continuous ranges of access. It splits the addresses into a first-level region table, a second-level flat page table with a larger page size (2 MB) and a third-level small page table with a conventional page size of 4 KB. 

To support the immense parallelism of PIM systems the \emph{Emu 1 architecture} introduces a special \emph{Partitioned Address Space} \cite{Minutoli2015}. With naming schemes it allows to map consecutive page addresses to interleaved PIM modules. As a result, the application is able to define the striping of data structures across all modules in the system.

\subsection{Data Coherence and Memory Consistency}
\label{sec:coherence}
Whenever multiple processors in disjoint coherence domains access the same shared data inconsistent states may occur due to missing cache coherence. Hence, cache coherence is crucial for preserving existing programming models and to PIM proliferation. Unfortunately, with increasing parallelism and number of coherence domains the burden of preserving memory consistency and cache coherence aggravated, to the point the where increased coherence traffic may cancel the improvements through PIM. Therefore, traditional fine-grained cache-coherence protocols implemented in modern multi-core CPUs (MESI, MESIF) are ill-suited for PIM settings.

One possible solution to this issue is to avoid on-chip caches in general or prevent caches to store data for a longer period of time than necessary. For instance, NDA's architecture does not provide the ability to access caches of the processor by the CGRAs \cite{Farmahini-Farahani2015}, just as data produced or modified by the processor should not be stored in its cache, but rather in a specific memory region, which is declared as un-cacheable. While CGRAs can consume data directly from this region, processors have to use non-temporal instructions (e.g., MOVNTQ, MOVNTPS, and MASKMOVQ in x86) that bypass the cache hierarchy. Due to this, whether processors nor CGRAs operate simultaneously on the same data and therefore avoid inconsistencies in the memory.

A clearly more flexible approach introduce specific cache operations. For instance, \cite{Gokhale2015} maintain memory consistency by executing cache flush and invalidate operations at well-defined synchronization points. These operations can either comprise the entire cache or specific address ranges in the granularity of cachelines. For example, during the communication of CPU and PIM module through shared memory, the sender must flush any modified data from its cache and the receiver must invalidate any associated address range. Further papers \cite{Hall1999,Draper2002,Nair2015} refer to the same simple coherence protocol to ensure consistency between the host cache and the PIM memories but have little differences in their implementation or their granularity (e.g. cacheline size, hardware/software implementations).
\cite{Ahn2015ISCA} manages improve on the above by knowing the exact cache blocks of a PIM instruction. Therefore it only has to issue invalidations or writebacks for these target cache blocks before processing the PIM operation. This should happen infrequently in practice since these operations are offloaded to the PIM memories with the expected data present on it. 

LazyPIM \cite{Boroumand:CAL:LazyPIM:2017} is an approach for efficient cache coherence in PIM settings, which overcomes the downsides of traditional cache-coherence protocols (MESI, MESIF) implemented in current multi-core CPUs. The basic idea behind LazyPIM is to allow speculative/optimistic PIM kernel execution, as if no cache-conflicts would occur and all permissions were granted. Upon successful execution all changed cache lines are transmitted to the host CPUs in a compressed form, where conflict detection in the CPUs cache coherence directory is performed. If no PIM cachelines conflict the CPU cache the PIM kernel successfully terminates. Otherwise its execution is rolled back, the CPU state is propagated to DRAM, and the PIM kernel is re-executed.

\subsection{Distributing and Partitioning Workload}
\label{sec:distribution}
Besides the previously described aspects, the distribution and partitioning of data and/or workload looms an important issue in PIM settings. Yet it is an NP-hard problem. The PIM performance improvements (parallelism and scalability, bandwidth utilization and low latencies) are only available when PIM kernels are executed on data directly located next to the processing unit. This is reminiscent of distributed systems, but involves only a single scale-up PIM system. 

A popular platform, making use of data distribution in a very large scale, is IBM's Netezza \cite{Francisco2011}. Its architecture is able to be extended by multiple intelligent processing nodes, called S-Blades. Equipped with multi-engine FPGAs, multi-core CPUs and gigabytes of memory they provide an excellent engine for massively parallel programs (MPP). Over a network fabric those S-Blades are connected to the host, which manages the data partitioning and query distribution. For instance, it compiles SQL queries into executable code segments and distributes these to the MPP nodes for execution. Important, yet, simple data structures for skipping partitions during processing are the so called \emph{Neteeza ZoneMaps}. BlueDBM \cite{Ming2015} follows a similar approach. Their nodes consist of Flash memory and a FPGA is responsible for in-storage processing and acts as interface controller for the various interfaces, e.g. flash or network. Thereby each node is a functional unit on itself and they can be connected in various network topologies (distributed star, mesh or fat tree).  

One of the first PIM systems investigating different distribution strategies is JAFAR \cite{Xi2015,Babarinsa2015}. JAFAR allows the user to decide how the address space is to be organized. Thereby, the system can be configured as a contiguous space where each DIMM is filled up after another or in an interleaved manner across multiple DIMMs. The latter expects symmetric DIMMs with the same capacity and latency.

\subsection{Summary}
Succinctly summarized, current work on computer architectural aspects of PIM are available across all levels of the memory and storage hierarchy as well as on todays concepts for modern multi-core systems such as address translation and cache coherence. However, these are only touched on their surface and require further investigation. Fairly new computational and programming models are suggested but there is still much room for improvement with regard to the full utilization of nowadays and tomorrows hardware capacities. Moreover, a number of novel ISAs are proposed for specific use cases while research about generic instruction sets for multiple purposes are still rare. In total, the current state of research is promising to exploit PIM against the challenges of Moore's Law and the end of Dennard scaling.

\section{Implications to Data Management and Processing}
The fundamental changes in nowadays hardware (Section \ref{sec:technologies}), and the accompanying effects on the concepts of modern computer architectures (Section \ref{sec:computerarchitecture}), have direct implications on data management and processing. These include data management operations such as scan, sort, group, join and index. The influence is also noticeable in the query evaluation in general, since the evaluation model needs to support the properties of the underlying hardware. Furthermore, atomicity of operations can be ensured within the hardware. As a result, new transactional protocols are proposed for PIM. The largest area of application for PIM research is currently the specific workloads. These reach from complex mathematical problems like discrete cosine transformations or nearest neighbor search to complex analytical algorithms such as clustering, graph processing or neural networks. The following sections present research in all of those categories (see Table \ref{tab:stateofresearch}).

\renewcommand{\arraystretch}{1.8}
\begin{table*}
	\scriptsize
	\caption{Chronologically ordered state of research including their storage and integration technology. For each approach the classification in Data Management and Computer Architecture of Table \ref{tab:classification} is given.}
	\label{tab:stateofresearch} 
	\begin{tabular}{llllp{1cm}p{1.5cm}p{3cm}p{3.5cm}}
		\hline
		\textbf{\#} & \textbf{Name} & \textbf{Ref.} & \textbf{Year} & \textbf{Storage} & \textbf{Integration} & \textbf{Data Management} & \textbf{Computer Architecture}\\
		\hline
		1 & EXECUBE & \cite{Kogge1994} & 1994 & DRAM & Circuit & \diamonded{8} & \circled{1}\\
		\hline
		2 & Terasys & \cite{Gokhale1995} & 1995 & SRAM & Circuit & \diamonded{10} & \circled{1} \circled{2} \circled{3}\\
		\hline
		3 & IRAM & \cite{Kozyrakis1997} & 1997 & DRAM & Circuit & \diamonded{10} &  \\
		\hline
		4 & Active Pages & \cite{Oskin1998} & 1998 & DRAM & Circuit & \diamonded{1} \diamonded{10} & \circled{1} \\
		\hline
		5 & FlexRAM & \cite{YiKang1999} & 1999 & DRAM & Circuit &  & \circled{3}\\
		\hline
		6 & Active Disks & \cite{Riedel2001,Riedel1999} & 1999 & HDD & PCB & \diamonded{10} & \circled{5}\\
		\hline
		7 & DIVA & \cite{Hall1999,Draper2002} & 1999 & DRAM & Circuit & \diamonded{5} & \circled{3} \\
		\hline
		8 & Computational RAM & \cite{Elliott1999,Elliott2008} & 1999 & DRAM & Circuit & \diamonded{8} & \circled{1} \\
		\hline
		9 & ASF & \cite{Fitch2009} & 2009 & Flash & PCB & \diamonded{1} & \circled{3} \\
		\hline
		10 & IBM Netezza & \cite{Francisco2011} & 2011 & Storage & Platform & \diamonded{6} & \circled{5} \\
		\hline
		11 & Minerva & \cite{De2013} & 2013 & DRAM & PCB & \diamonded{6} & \circled{1} \\
		\hline
		12 & Active Flash & \cite{Tiwari:ActiveDiskInSitu:FAST:2013} & 2013 & Flash & PCB & \diamonded{9} & \\
		\hline
		13 & iSSD & \cite{Cho2013} & 2013 & Flash & PCB & \diamonded{1} & \circled{1} \\
		\hline
		14 & 3D Sparse matrix mul. & \cite{Zhu2013HPEC,Zhu2013} & 2013 & DRAM & Package & \diamonded{10} &\\
		\hline
		15 & SmartSSD & \cite{Kang2013} & 2013 & Flash & PCB & \diamonded{7} & \circled{2} \\
		\hline
		16 & IBEX & \cite{Teubner:IBEX:SIGMOD:2013,Teubner:IBEX:VLDB:2014} & 2014 & Flash & Platform & \diamonded{4} & \\
		\hline
		17 & Willow & \cite{Seshadri2014} & 2014 & Flash & PCB & \diamonded{6} & \circled{2} \\
		\hline
		18 & NDC & \cite{Pugsley2014} & 2014 & DRAM & Package & \diamonded{7} & \\
		\hline
		19 & TOP-PIM & \cite{Zhang2014} & 2014 & DRAM & Package & \diamonded{9} & \circled{1} \\
		\hline
		20 & Q100 & \cite{Wu2014} & 2014 & DRAM & PCB & \diamonded{6} & \circled{2}\\
		\hline
		21 & JAFAR & \cite{Xi2015,Babarinsa2015} & 2015 & DRAM & PCB & \diamonded{1} & \circled{2} \circled{3}\\
		\hline
		22 & TESSERACT & \cite{Ahn2015} & 2015 & DRAM & Package & \diamonded{9} & \circled{1} \\
		\hline
		23 & DRE & \cite{Gokhale2015} & 2015 & DRAM & PCB & \diamonded{9} & \circled{3} \circled{4} \\
		\hline
		24 & Bitwise AND and OR & \cite{Seshadri2015} & 2015 & DRAM & Circuit & \diamonded{3} & \circled{2}	\\
		\hline
		25 & Radix Sort on Emu 1 & \cite{Minutoli2015} & 2015 & DRAM & PCB & \diamonded{2} & \circled{1} \circled{3} \circled{5}	\\
		\hline
		26 & ProPRAM & \cite{Wang2015} & 2015 & PCM & Packaging & \diamonded{9} &  \circled{1} \circled{3}\\
		\hline
		27 & BlueDBM & \cite{Ming2015} & 2015 & Flash & PCB & \diamonded{10} & \circled{5} \\
		\hline
		28 & NDA & \cite{Farmahini-Farahani2015} & 2015 & DRAM & Package & \diamonded{9} & \circled{1} \circled{4} \\
		\hline
		29 & PIM-enabled & \cite{Ahn2015ISCA} & 2015 & DRAM & Package & \diamonded{5} \diamonded{10} \diamonded{11} & \circled{2} \circled{3} \circled{4}\\
		\hline
		30 & AMC & \cite{Nair2015} & 2015 & DRAM & PCB &  & \circled{1} \circled{2} \circled{3} \circled{4} \\
		\hline
		31 & Sort vs. Hash & \cite{Mirzadeh2015} & 2015 & DRAM & Package & \diamonded{5} & \\
		\hline
		32 & HRL & \cite{Gao2016} & 2016 & DRAM & Package & \diamonded{6} \diamonded{10} \diamonded{11} & \\
		\hline
		33 & SSDLists & \cite{Wang2016} & 2016 & Flash & PCB & \diamonded{1} \diamonded{2} & \\
		\hline
		34 & IMPICA & \cite{Hsieh2016} & 2016 & DRAM & Package & \diamonded{10} & \circled{3} \\
		\hline
		35 & TOM & \cite{Hsieh2016ISCA} & 2016 & DRAM & Package & \diamonded{9} & \circled{2} \circled{5}\\
		\hline
		36 & BISCUIT & \cite{Gu2016} & 2016 & Flash & PCB & \diamonded{1} & \circled{1} \circled{3}\\
		\hline
		37 & Tetris & \cite{Gao2017} & 2017 & DRAM & Package & \diamonded{11} & \\
		\hline
		38 & CARIBOU & \cite{Ist2017} & 2017 & DRAM & PCB & \diamonded{1} & \circled{3}\\
		\hline
		39 & Sorting big data & \cite{Vermij2017} & 2017 & DRAM & Package & \diamonded{2} & \\
		\hline
		40 & SUMMARIZER & \cite{Koo2017} & 2017 & Flash & PCB &  \diamonded{1} & \circled{2}  \\
		\hline
		41 & MONDRIAN & \cite{Drumond2017} & 2017 & DRAM & Package & \diamonded{1} \diamonded{2} \diamonded{4} \diamonded{5} & \circled{1} \\
		\hline
		42 & LazyPIM & \cite{Boroumand:CAL:LazyPIM:2017} & 2017 & DRAM & Package & \diamonded{10} & \circled{4} \\
		\hline
		43 & CAIRO & \cite{Hadidi2017} & 2017 & DRAM & Package & & \circled{2} \\
		\hline
		44 & XMem & \cite{Vijaykumar2018} & 2018 & DRAM & Simulator &  & \circled{1} \circled{2} \circled{3} \circled{4}\\
		\hline
		45 & PIM for NN & \cite{Liu2018} & 2018 & DRAM & Package & \diamonded{11} & \circled{1}\\
		\hline
	\end{tabular}
\end{table*}

\renewcommand{\arraystretch}{1.7}
\begin{table*}
	\scriptsize
	\caption{Classification of PIM approaches in Data Management and Computer Architecture Categories}
	\label{tab:classification} 
	\begin{tabular}{clp{5.4cm}clp{5.2cm}}
		\hline
		\textbf{Symbol} & \textbf{Section} & \textbf{Data Management} 	& \textbf{Symbol} & \textbf{Section} & \textbf{Computer Architecture}\\
		\hline
		\diamonded{1} & \ref{sec:scanning} & Scanning/Filtering						& \circled{1} & \ref{sec:computationalmodel} & Computational/Programming Model \\
		\diamonded{2} & \ref{sec:sorting} & Sorting									& \circled{2} & \ref{sec:instructionset} & Instruction Set \\
		\diamonded{3} & \ref{sec:indexing} & Indexing								& \circled{3} & \ref{sec:memorymanagement} & Addressing \\
		\diamonded{4} & \ref{sec:grouping} & Grouping								& \circled{4} & \ref{sec:coherence} & Coherence \\
		\diamonded{5} & \ref{sec:joining} & Joining									& \circled{5} & \ref{sec:distribution} & Distributing/Partitioning \\
		\diamonded{6} & \ref{sec:queryeval} & Query Evaluation						&  & &\\
		\diamonded{7} & \ref{sec:distributedprocessing} & Distributed Processing	 				&  & &\\
		\diamonded{8} & \ref{sec:dct} & Discrete Cosine Transform 				&  & &\\
		\diamonded{9} & \ref{sec:clustering} & Clustering								&  & &\\
		\diamonded{10} & \ref{sec:graphprocessing} & Graph/Matrix Processing				&  & &\\
		\diamonded{11} & \ref{sec:nn} & Neural Networks and Deep Learning		&  & &\\
		\hline
	\end{tabular}
\end{table*}

\subsubsection{Scanning and Filtering}
\label{sec:scanning}
One of the prominent and basic operations in data management systems are scans. They are highly data intensive, since each value have to be accessed, whenever the dataset is not pre-sorted, and compared to the filtering conditions. Therefore, scans require fast storage-/memory-processing interaction, which is one of the declared goals of PIM and is addressed in various research. Well-known scan optimizations form the field of main-memory DBMS like \emph{SIMD-scan} \cite{Willhalm:SIMDScan:VLDB:2009} or \emph{BitWeaving} \cite{Li:BitWeaving:SIGMOD:2013} can benefit significantly from PIM.

Already in 1998, Active Pages \cite{Oskin1998} investigated that operations on a page-level basis are required to address various workloads for PIM. With their flexible interface, they are able to bind and execute different operations on the PIM modules. These are build up on RADram, an integration of FPGAs and DRAM technology, which can exploit an extremely high parallelism. Applied on database queries the authors claim to speed up searches of un-indexed datasets over 10 times.

A research group of IBM propose the Active Storage Fabrics (ASF) \cite{Fitch2009} to tackle petascale data intensive challenges. ASF lays between the host workloads (e.g. TPC-H) and the Blue Gene Compute Nodes. The central component is a Parallel in-Memory Database (PIMD), which stores KV-Pairs within Partitioned Data Sets distributed across those nodes. \textit{Parallel Data Intensive Primitives}, such as scans, are executed on the ASF layer and are distributed over the entire nodes to leverage the full parallelism. 

\begin{figure}
	\includegraphics[scale=0.3]{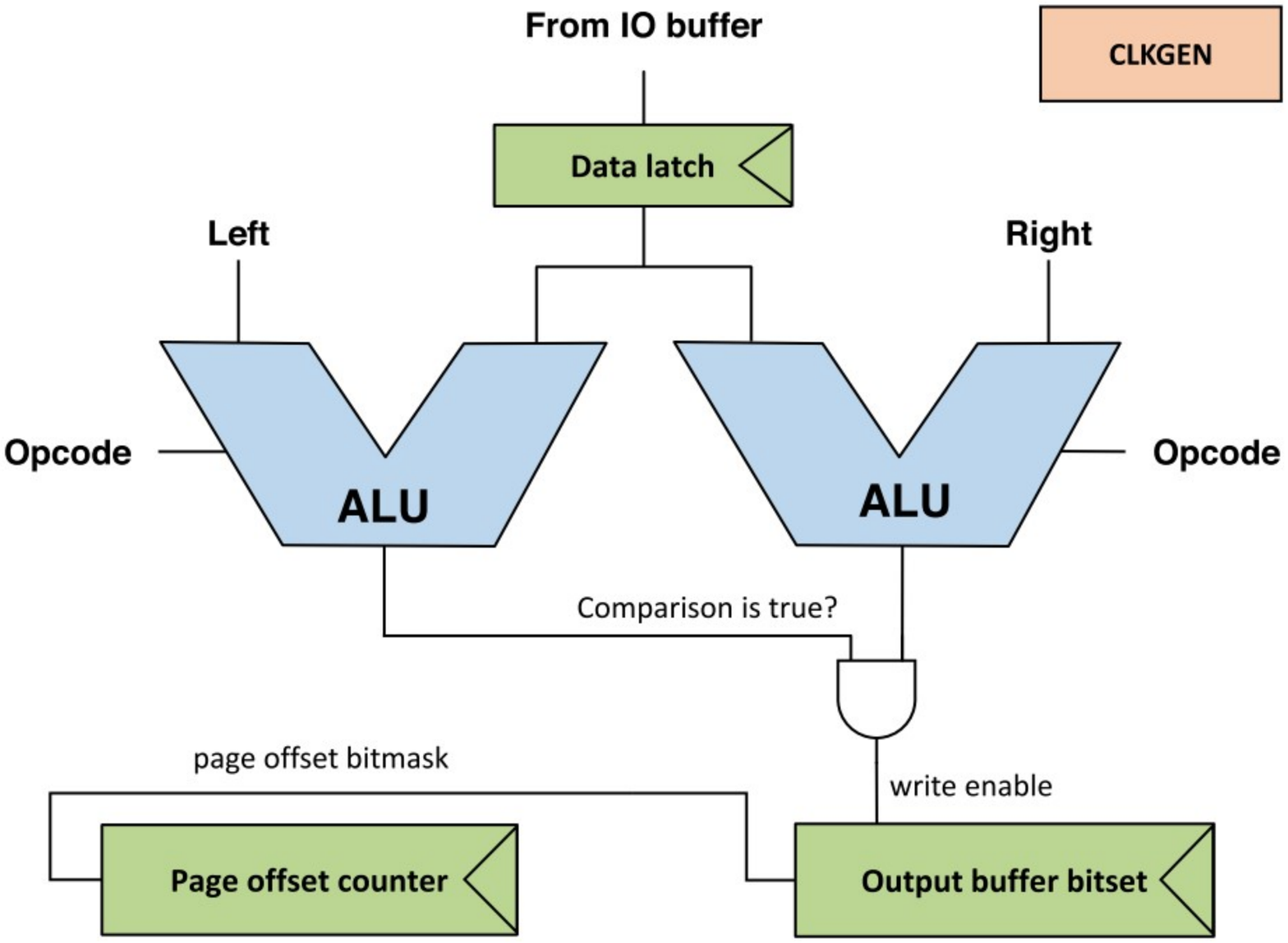}
	\caption{JAFAR's Architecture Diagram (from \cite{Xi2015}).}
	\vspace{-5pt}
	\label{fig:JAFAR} 	
\end{figure}

JAFAR \cite{Xi2015,Babarinsa2015} is a column-store accelerator, designed by the university of Harvard, to offload selects to the memory as PIM kernels, gaining a performance improvement of about 9x. It supports classical comparison predicates (=, $<$, $>$, $\leq$, $\geq$) applied on values of the data type integer. Its architecture, shown in Figure \ref{fig:JAFAR}, comprises two ALUs that can work in parallel to enable range filters. Whenever JAFAR's API is called with a pointer to a virtual memory start address, the JAFAR hardware starts issuing the respective read requests against the DRAM modules. Each received 64 bit word is processed by the ALUs. The result is a bitmask, indicating, which rows passed the filter operands, written back to the DRAM and memory mapped by the host system. By polling a shared-memory location the host is informed about the completion.

A similar approach is proposed with the iSSD \cite{Cho2013}. Here, the flash memory controller is extended by a stream processor comprising an array of ALUs. These could either be pipelined to compute higher-order functions or be connected to the resident SRAM to store temporary results. A configuration memory enables to change the data flow during runtime. In the evaluation, the authors show a 2.3x improvement when scanning a 1 GB TPC-H Lineitem dataset with a selectivity of 1\% in contrast to the standard device-to-host communication.

\begin{figure}
	\includegraphics[scale=0.3]{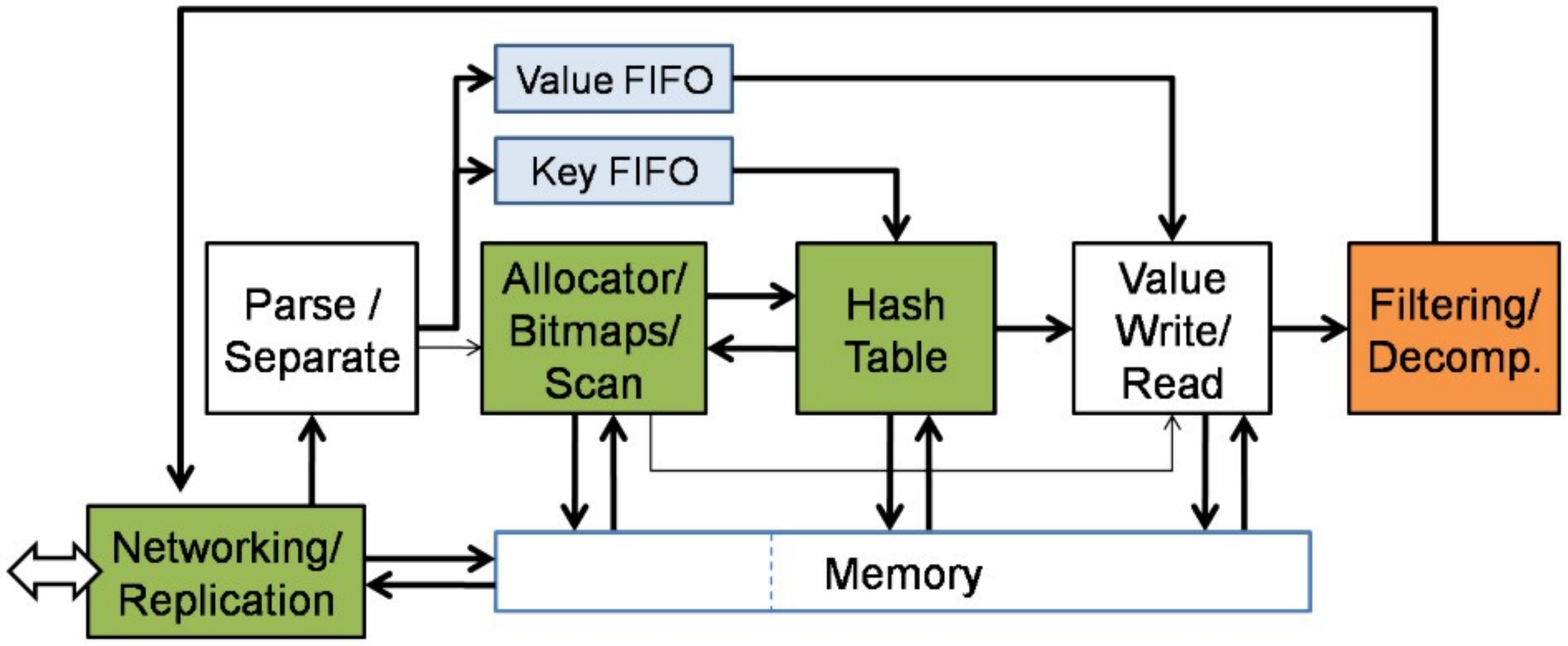}
	\caption{Caribou's Architecture Diagram (from \cite{Ist2017}).}
	\vspace{-5pt}
	\label{fig:caribou} 	
\end{figure}

Caribou \cite{Ist2017} implements a slightly different approach with a distributed key-value store, which persists the primary key as a key and the remaining fields as a value. This shared-data model is replicated from the master to the respective replica nodes using Zookeeper's atomic broadcast. The nodes are connected to a conventional 10 Gb/s switch and equipped with Xilinx Virtex 7 FPGAs and 8 GB of memory. The block diagram of Figure \ref{fig:caribou} shows the most important components of one module, whereby the Allocator/Bitmap/Scan area is responsible for the scans. By constantly managing two bitmaps, one for the allocated memory and one for the invalidated addresses, the scan module is able to issue a read command to memory for each bit set to 1. Whenever data is fetched, a pipelined comparator can be used to filter the keys or values on a specific selection predicate. The performance is limited by either the selection itself (low selectivity) or the network (high selectivity).

Another system implemented the scan operation is the Summarizer \cite{Koo2017}. Like \cite{Cho2013}, it uses the processing unit near the flash modules to implement a task controller. It is attached to the traditional SSD controller for FTL and I/O command management as well as to the flash and DRAM controller. By extending the NVMe command set with new commands the Summarizer is able to execute user defined functions such as filtering upon a traditional READ command. 

\subsubsection{Sorting}
\label{sec:sorting}
Another data-intensive operation is \emph{Sort}. \cite{Minutoli2015} use the Emu 1 system to implement basic sorting in a PIM fashion. The Emu1 system consists of multiple nodes, which are divided into Stationary Cores, Nodelets, NVRAM, and a Migration Engine, which handles the interconnection of the nodes. While, the Stationary Cores implement a conventional ISA and thus run classical operating systems, the \emph{Nodelets} are the basic building block for near-memory communication. These comprise a Queue Manager, a local Thread Store, a special Gossamer Core and Narrow Channel DRAM. Its idiosyncrasy is to migrate lightweight threads from one \emph{Nodelet} to another and thereby avoid remotely loading data from one core to another. In their evaluation, they apply this advantage to a radix sort, which usually partitions the initial dataset of N keys into M blocks and to sort in parallel. In the Emu 1 system, each block is assigned to a thread computing a local histogram. Later, these histograms are merged to a global histogram and, upon this, the offsets for groups of the same key can be calculated. However, the performance decreases with around 32 threads because of the high effort of migrations in contrast to the actual calculations, known from other research as well \cite{Cho2013}.

Another system for PIM-sorting is the Mondarian Data Engine \cite{Drumond2017}. Like \cite{Minutoli2015}, its a distributed system of multiple PIM devices connected within a network and managed by the host using a conventional CPU. The devices build up on Intel and Micron's HMC technology \cite{Pawlowski2011}. However, based on their first-order analysis, the authors conclude that it is difficult to saturate the internal bandwidth solely with conventional  MIMD cores and introduced streaming buffers to avoid any memory access stalls. Along the lines of their analytical operators analysis, the authors identified sorting as a major research topic for such data streams. To leverage to full potential, they use the data permutability property to convert random access patters into sequential ones. Thereby, they conclude that algorithms for CPUs do not fit properly for modern PIM execution and have to be radically adapted.

Same holds for \cite{Vermij2017}, which analyses a merge sort execution on multiple PIM devices. As they implement their algorithm on a reconfigurable fabric, they developed a workload-optimized merge core in VHDL to do a single partial merge. This is necessary because the latest iterations of the merge sort algorithm involve larger data sets. In contrast to the first iterations, where data sets are small, these merge executions cannot be parallelized in a straightforward way and require a special implementation to leverage the full potential of the PIM devices. 

\subsubsection{Indexing}
\label{sec:indexing}
Indexing is a classical technique to improve query performance, yet index operations cause data transfers (lookup) and transfer overhead (pointer chasing, maintenance, e.g. sorting, just to name a few sources). Index operations are mainly based on search key comparisons. As those operations seem to be predestined for PIM a few research works focus on either the comparison or the management of index structures.

\begin{figure}
	\includegraphics[scale=0.3]{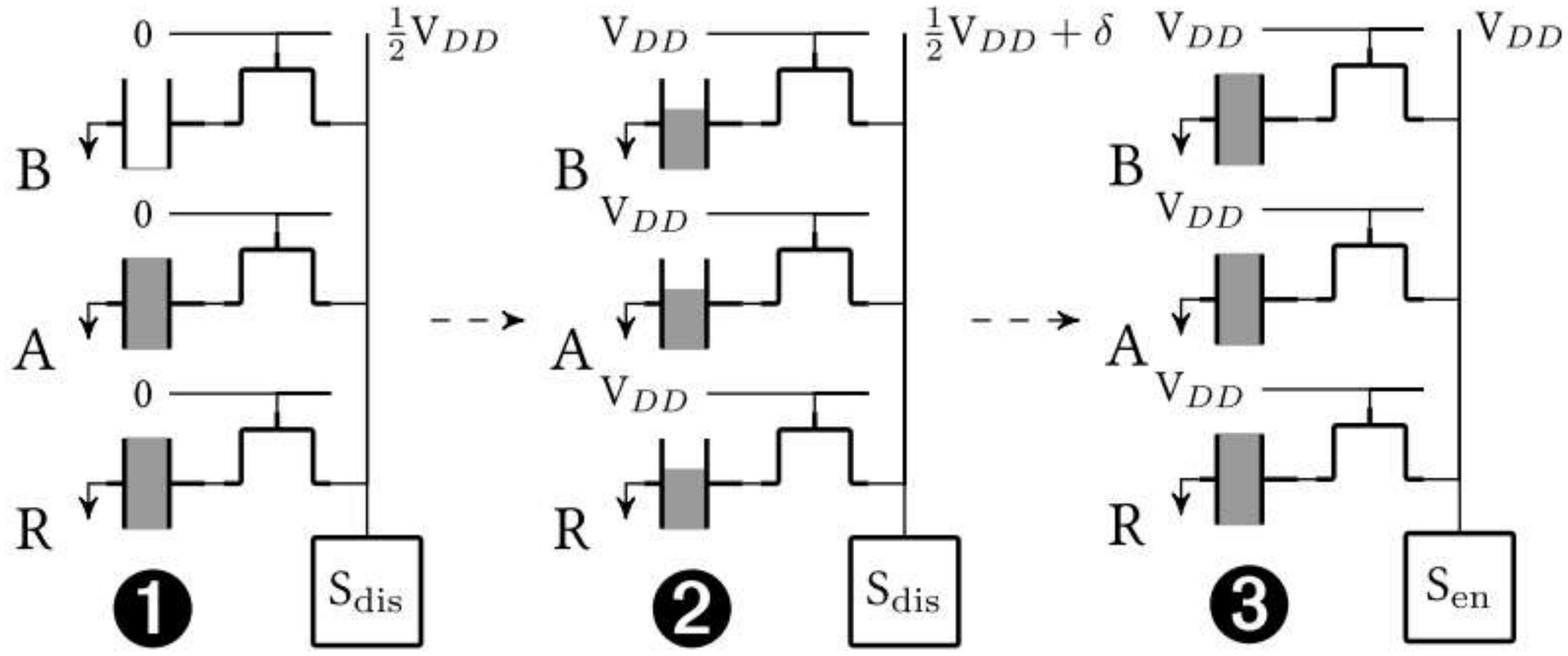}
	\caption{Example for bulk bitwise AND and OR (from \cite{Seshadri2015}).}
	\vspace{-5pt}
	\label{fig:bitwise} 	
\end{figure}

For instance, a research group of the Carnegie Mellon University in cooperation with Intel recognized that fast bulk bitwise AND and OR operations are important components of modern day programming \cite{Seshadri2015} and especially in bitmap indexes, which are very widespread in analytical (business intelligence) DBMS. The primitives can be implemented directly within the fabric logic of DRAM. When simultaneously connecting three cells to a bitline, their resulting bitline voltage after charge sharing is equivalent to the majority value of these three cells. Consider the example shown in Figure \ref{fig:bitwise}. Firstly, two of the three cells are positively charged. Secondly, after connecting, there is a positive deviation on the bitline voltage, letting, thirdly, all cells become fully charged. Expressed in logic, this phenomenon complies to $R(A+B) + \bar{R}(AB)$, which offers to switch between AND and OR using the state of R. Their evaluation show an improvement of 9.7x higher throughput and 50.5x lower energy consumption compared to standard vector processing \cite{IntelAVX}, which have to read all the data from DRAM to the CPU.

Other widespread index structures, like B/B+-trees, are limited in its performance due to pointer chasing. \cite{Hsieh2016} tries to minimize the effects by performing pointer chasing operations directly inside the memory utilizing PIM. Thereby, they claim to reduce the latency of such operations, since addresses do not have to be transferred to the CPU, and ease the caches, which are inefficient for pointer chasing. This is achieved by several new techniques within the computer architecture, like the region-based page table, described in section \ref{sec:memorymanagement}. Similarly, \cite{Hong:ALT:PACT:2016} describes and NDP approach to \emph{linked-list traversal} on secondary storage, whereas \cite{Wang:SSDLists:DAMON:2016} tackles the problem of \emph{NDP list intersection}.

\subsubsection{Grouping}
\label{sec:grouping}
There is also some promising research implementing grouping operations of databases on FPGAs as accelerators. Since reconfigurable fabrics are often part of PIM devices, these investigations could be seen as ground work from the database instead of the computer architecture perspective. 

A  prototype for an intelligent storage engine called IBEX is proposed in \cite{Teubner:IBEX:SIGMOD:2013,Teubner:IBEX:VLDB:2014}. Besides various other database operations, grouping is implemented on a Vertex 4 FPGA. Utilizing a hash table, keys are compared to the selection criteria and, in case of a match, the respective values are directly aggregated in a pipelined fashion. Thereby, the input is 256 bit wide and can be flexibly divided into multiple keys and value to support combined group keys. Within their evaluation, they show dramatic throughput improvements of Ibex over two standard storage engines, MyISAM and INNODB. 

\subsubsection{Joining}
\label{sec:joining}
Joins represent a performance critical task in todays transactional and analytical queries on large datasets. Independent on the join type, these operations have to compare all values of the involved relations with respect to the join attributes. Therefore, it is heavily data intensive, offer potential for efficient parallelization. In contrast to other operations, which are size-reducing (the result size is smaller than the input dataset size), this property cannot be guaranteed for joins. Hence, under certain conditions (e.g. data distribution and join condition), joins may amplify data transfers, which is a major pitfall. 

DIVE \cite{Hall1999,Draper2002}, the Data Intensive Architecture, is proposed by Mary Hall et al. Within their evaluation they demonstrate the potential of the PIM-based architecture on several application and algorithms, inter alia, a natural-join. Therefore, they build up hash tables for both relations with an index on the given attribute. The algorithm joins every tuple of the two relations that share a common value. This happens for every PIM node by firstly distributing a set of consecutive entries of the hash table; secondly, computing a local natural join; and thirdly, merging the partial hash tables to gain the result.

A similar approach is proposed by \cite{Ahn2015ISCA}. For their case study of the PIM-enabled instructions they support a hash join of an in-memory database. Because this join builds a hash table with one relation and probes it with keys of the other, it requires an efficient PIM operation for hash table probing. While the PIM device compares the keys in a given bucket, the host processor issues the PIM operations and merges the results. Since their implementation allow multiple hash table lookups for different rows to be interleaved, multiple PIM operations can be triggered as an out-of-order execution.

\cite{Mirzadeh2015} evaluate the differences between the common sort and hash join algorithms with focus on near-memory execution on HMC. Thereby, they improved the performance and energy-efficiency by carefully considering the data locality, access granularity and microarchitecture of the stacked memory.

\cite{Teich:FPGAsql:TRTS:2016} proposes a further approach for an FPGA implementation of different join algorithms suitable for NDP.

\subsection{Query Evaluation}
\label{sec:queryeval}
The adoption to PIM is not only limited to low storage functions as described in the previous sections, but rather can improve performance on query evaluation level. Research is currently spread across NDP scenarios, but can easily adapted to PIM. 

One popular system makes use of PIM's prevention of unnecessary data transfers, by pushing down processing to the data, is IBM's Netezza \cite{Francisco2011}. The system includes FPGAs in the disk controller, which are able to execute parts of queries. The partial results of such local processing are sent back and merged by the management unit. 

\begin{figure}
	\includegraphics[scale=0.3]{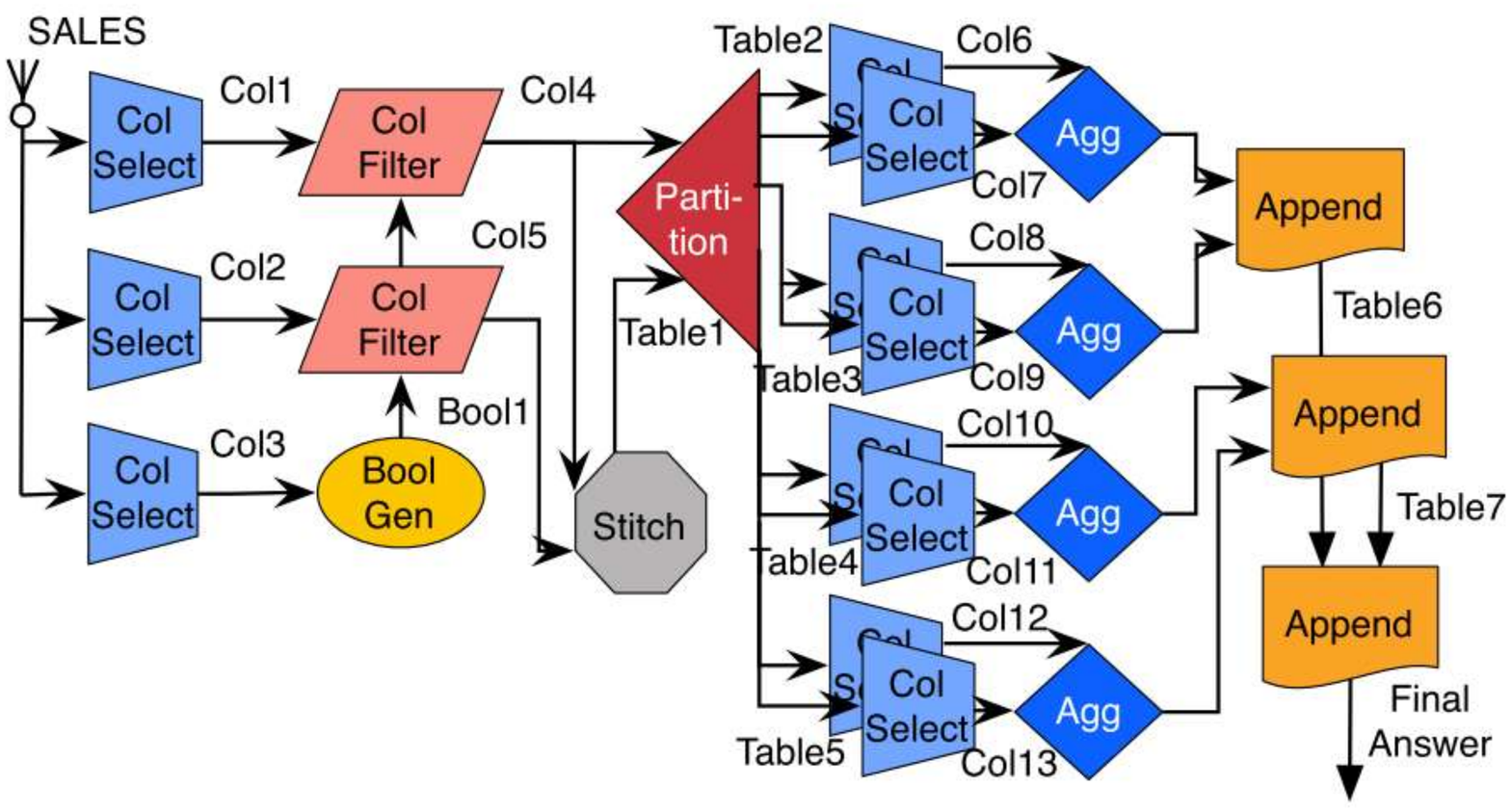}
	\caption{Example query transformed into Q100 spatial instructions (from \cite{Wu2014}).}
	\vspace{-5pt}
	\label{fig:q100} 	
\end{figure}

Other research works, like Q100 \cite{Wu2014}, present accelerators specifically designed for database processing. A collection of heterogeneous ASICs is able to process relational tables and columns efficiently in terms of throughput and energy. Data streamed through these ASICs are manipulated using a coarse grained instruction set comprising all standard relational operators. Figure \ref{fig:q100} shows an exemplary query plan transformed into the Q100 spatial instructions. Depending on the available resources, this graph is broken into temporal instructions, which are executed sequentially. This exploits the full potential of pipelining.

Often research focuses on less complicated database management applications like KV-Stores. As their API includes only simple Put and Get instructions there is no complex query evaluation, but their throughput and latency are of high importance. Therefore, Minerva \cite{De2013} tries to use their FPGA based system to accelerate KV-Stores by reducing data traffic between the host and storage. It allows to offload data intensive tasks, like searching of specific key patterns, to the NVM storage controller. Thereby, it performs up to 5.2M get and 4.0M put operations/s, which is about 7-10 times more than a conventional PCM-based SSD. 

Besides reducing data transfers, further opportunities are possible by PIM. For instance, it is possible to change the limits of properties like atomicity. \cite{Seshadri2014} purposes Willow, a user-programmable SSD, able to execute application logic on the storage device similar to a remote procedure call. Thereby, one of the demonstrated case studies is the execution of atomic writes. Atomic writes are very well known in database management system to enforce consistency, e.g. through write-ahead logging (WAL). They occur in simple journaling mechanisms but also in complex transaction protocols like ARIES. However, with the new characteristics of PIM even new protocols are possible. To this end, MARS \cite{Coburn2013} is a novel WAL scheme with the same functionality like ARIES, but without the disk-centric implementation decisions and thereby, revise the transactional semantics of PIM-enabled databases. In their evaluation, the authors show an improvement of up to 1.5x of MARS over traditional ARIES with Direct IO \cite{Seshadri2014}.

\subsection{Domain-specific Operations}
\label{sec:workload}
Instead of flexible instruction sets like database operators, often application specific algorithms are part from PIM research in the data management area. These range from mathematical problems (e.g BLAS), which have a high execution complexity, to distributed processing. 

\subsubsection{Distributed Processing}
\label{sec:distributedprocessing}
Modern workloads (e.g. HTAP) and analytical operations (e.g. processing large graphs) are often too large to process by a single server. As a result, problems are broken up, distributed on multiple instances, and combined to a final result afterwards. The probably most famous algorithm for such compute scenarios is nowadays MapReduce. Various research have taken PIM approaches to improve both the map and the reduce phases of the model.

For instance, the SmartSSD \cite{Kang2013} allows to create on-device map functions, which are called after splitting up the input files. The parameters involve a range of logical addresses (object IDs) to identify the respective data. It then performs a logical combine and reduce phase. Only the results are communicated to the host system to minimize disk traffic.

A similar approach is presented with NDC \cite{Pugsley2014}. Instead of flash and in-storage processing they focus on real in-memory processing by utilizing HMC as base technology. The user provides map and reduce functions, which are transparently executed by the NDC cores located on the reconfigurable fabrics of the HMC. Each of these cores is associated with a vertical memory slice of 256 MB that is likewise the data layout for NDC applications. By overcoming the bandwidth wall \cite{Burger1996} with this setup they can reduce the execution time by considerable 12.3\% to 93.2\%.

\begin{figure}
	\includegraphics[scale=0.3]{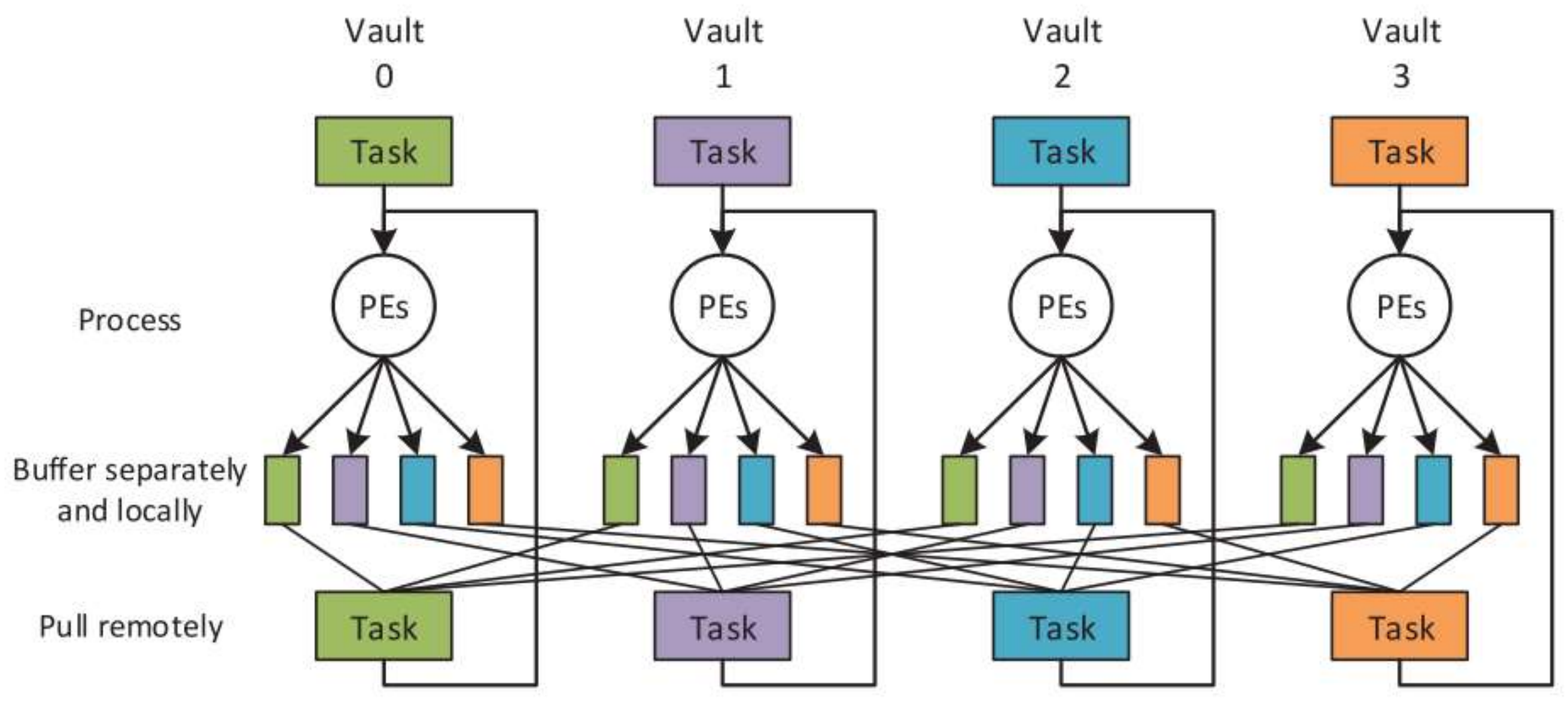}
	\caption{Example execution distribution of HRL (from \cite{Gao2016}).}
	\vspace{-5pt}
	\label{fig:hrl} 	
\end{figure}

Heterogeneous Reconfigurable Logic (HRL) \cite{Gao2016} represents a more flexible approach. Utilizing FPGAs or CGRA arrays, HRL provides coarse-grained and fine-grained logic blocks, separates routing networks for data and control signals, and includes specialized units for branch operations and irregular data layouts. Its execution flow is fairly simple and generic. Each processing element start in parallel to process its assigned data while holding the results in its own local buffers as shown in Figure \ref{fig:hrl}. When all data is processed, the host is notified and the next iteration is started.

\subsubsection{Discrete Cosine Transformation}
\label{sec:dct}
In the mid and late 1990s the semiconductor industry was able to fabricate first versions of PIM modules on a 2D integrated circuit. The processing elements based mainly on consecutive linked logical elements rather than a processor with large instruction sets. Therefore, the problem space was limited to problems solvable by these logical gathers. However, since these elements were located directly within the circuit they could exploit the full bandwidth of the memory modules. As a consequence, Discrete Cosine Transformation (DCT) became a wide spread issue to solve with PIM modules to improve image processing, e.g. JPEG compression. A few representatives are the Computational RAM \cite{Elliott1999,Elliott2008} and the EXECUBE \cite{Kogge1994}.

\subsubsection{Clustering}
\label{sec:clustering}
Clustering is necessary to detect certain groups with similar properties. Current clustering algorithms like \emph{k-means} involve a lot of data, which have to be processed multiple times. With conventional hardware this leads to an immense amount of traffic on the buses, which is the reason why PIM is a favorable method to tackle this workload. As a consequence various research \cite{Hsieh2016ISCA,Farmahini-Farahani2015,Wang2015,Zhang2014,Tiwari:ActiveDiskInSitu:FAST:2013} evaluate their implemented hardware and software against data mining workloads. However, usually the design of the proposed approaches is clearly more flexible than just a simple clustering algorithm and focus mainly on computer architectural improvements as described in Section \ref{sec:computerarchitecture}.

\subsubsection{Graph Processing}
\label{sec:graphprocessing}
Graph analytics is a major research topic as the demand in the industry rises with increasing data volume. Since some of the algorithms traverse the graph multiple times it is a desirable application for PIM. Furthermore, graph processing matches PIM since it is also latency-bound. 

\cite{Zhu2013HPEC,Zhu2013} introduce a 3D stacked logic in memory (LiM) system to process sparse matrix data. Building a content addressable memory hardware structure, it is able to exploit the sparse data patterns for executing generalized sparse matrix-matrix multiplication without any software approach based techniques like heaps. Their simulation demonstrates more than two orders of magnitude of performance and energy efficiency improvements compared with the traditional multi-threaded software implementation on modern processors.

Another PIM accelerator for large-scale graph processing is TESSERACT \cite{Ahn2015}. Besides graph processing, TESSERACT focuses on fully utilizing the entire memory bandwidth and the communication of memory partitions. The proposed programming interface tries on the one side to exploit the hardware, and on the other side to improve graph processing by allowing the user to give hints about memory access characteristics like possible prefetches.

Additional research on nearest neighbor search \cite{Riedel2001,Riedel1999,Ming2015,Gokhale1995} exist, which is an operation frequently used in combination with graph processing. All those systems have in common that data transfers are reduced by partially offloading the execution of application code to the processing capabilities of the PIM modules.

A completely different approach is shown by the Data Rearrangement Engine (DRE) \cite{Gokhale2015}, which focuses on rearranging the hardware memory structures to dynamically restructure in-memory data to a cache-friendly layout and to minimize wasted memory bandwidth. The DRE consists of three instructions; setup, fill, and drain. The \textit{setup} loads parameter such as base addresses and payload sizes, the \textit{fill} copies the data from DRAM to the according buffers in the given pattern of \textit{setup}, and the drain copies the data from the buffers back into DRAM. This technique can be applied to graph processing applications like PageRank by repeatedly executing the setup and fill commands for each vertex with a minimum number of edges. Thereby, the \textit{fill} operation copies the list of the vertex's edges into the DRAM where the CPU can calculate the page rank accordingly.

\subsubsection{Neural Network and Deep Learning}
\label{sec:nn}
Modern deep neural networks emerged new accelerators to improve performance. However, these are not fit for the increasingly larger networks as they exceed on-chip SRAM buffers and off-chip DRAM channels. Tackling this issue, the presented scalable neural network (NN) accelerator Tetris \cite{Gao2017} leverages the high throughput and low energy consumption of 3D memory to increase the area of processing elements and reduce the area of SRAM buffers. Furthermore, Tetris moves NN computations partially to the DRAM and eases the contention on the buses. This leads to an improved data flow for NN accelerators. 

Similar issues are evaluated in \cite{Liu2018}. Due to the high data movement from memory to CPU during the training of NNs, PIM especially fits these requirements. The authors propose a heterogeneous co-design of hard- and software on basis of ARM cores and 3D stacked memory to schedule various NN training operations. To support program maintenance across the heterogeneous system, the OpenCL programming model is extended.

\subsection{Summary}
One can clearly tell that numerous parts of PIM concepts are already investigated in a large variety of data management applications. Most of them focus on specific workload-based problems like distributed processing, clustering, graph processing or neural networks as shown in Section \ref{sec:workload}. However, often these approaches can perfectly handle the given problem space but do not cover most of PIM's application variety in data management systems. For example, only a few papers currently investigate the implementation of classical database operators as PIM instructions or primitives. Yet, their research include many considerations about pipelining, executions semantics and atomicity. In sum, like in the general computer architecture of Section \ref{sec:computerarchitecture}, there is some promising work in many aspects of data management but cannot make any claim to be exhaustive.

\begin{acknowledgements}
This work has been supported by the project grant \emph{HAW Promotion} of the \emph{Ministry of culture youth and sports, state of Baden-W\"urrtemberg}, Germany.
\end{acknowledgements}

\bibliographystyle{spmpsci}      
\bibliography{pim_survey}

\begin{thebibliography}{100}
\providecommand{\url}[1]{{#1}}
\providecommand{\urlprefix}{URL }
\expandafter\ifx\csname urlstyle\endcsname\relax
  \providecommand{\doi}[1]{DOI~\discretionary{}{}{}#1}\else
  \providecommand{\doi}{DOI~\discretionary{}{}{}\begingroup
  \urlstyle{rm}\Url}\fi

\bibitem{3DXPoint}
Intel 3d xpoint.
\newblock http://www.micron.com

\bibitem{IntelAVX}
Intel isa extensions avx

\bibitem{RISCV}
Risv-v.
\newblock https://riscv.org/

\bibitem{ICNobelprize}
https://www.nobelprize.org/prizes/physics/2000/ (2000)

\bibitem{ITRS2011}
International Technology Roadmap Semiconductors  (2011)

\bibitem{Acharya:ASPLOS:ActiveDisc:1998}
Acharya, A., Uysal, M., Saltz, J.: Active disks: Programming model, algorithms
  and evaluation.
\newblock In: Proc. ASPLOS (1998)

\bibitem{Agerwala:DataIntensive:ICPP:2014}
Agerwala, T., Perrone, M.: Data centric systems: The next paradigm in
  computing.
\newblock In: Proc. ICPP (2014)

\bibitem{Ahn2015}
Ahn, J., et~al.: {A scalable processing-in-memory accelerator for parallel
  graph processing}  (2015)

\bibitem{Ahn2015ISCA}
Ahn, J., et~al.: {PIM-enabled instructions: A Low-Overhead, Locality-Aware
  Processing-in-Memory Architecture}.
\newblock Proc. ISCA  (2015)

\bibitem{Babarinsa2015}
Babarinsa, O.: {JAFAR : Near-Data Processing for Databases}.
\newblock Sigmod  (2015)

\bibitem{Balasubramonian:MSSC:MakingTheCaseForFeatureRichMemorySystems:2016}
Balasubramonian, R.: {Making the Case for Feature-Rich Memory Systems: The
  March Toward Specialized Systems}.
\newblock IEEE Solid-State Circuits Mag.  (2016)

\bibitem{Swanson:WONDP:Micro:2014}
Balasubramonian, R., et~al.: Near-data processing: Insights from a micro-46
  workshop.
\newblock IEEE Micro  (2014)

\bibitem{Binnig:DS:SmartNEtworks:2018}
Binnig, C.: Scalable data management on modern networks.
\newblock Datenbank Spektrum  (2018)

\bibitem{Boral:DatabaseMachines:1989}
Boral, H., DeWitt, D.J.: Parallel architectures for database systems.
\newblock chap. Database Machines: An Idea Whose Time Has Passed? A Critique of
  the Future of Database Machines (1989)

\bibitem{Borkar2010}
Borkar, S.: {3D integration technology for energy efficient system design}.
\newblock In: Proc. DAC (2010)

\bibitem{Boroumand2018}
Boroumand, A., Ranganathan, P., Mutlu, O., Ghose, S., Kim, Y., Ausavarungnirun,
  R., Shiu, E., Thakur, R., Kim, D., Kuusela, A., Knies, A.: {Google Workloads
  for Consumer Devices}.
\newblock ACM SIGPLAN  (2018)

\bibitem{Boroumand:CAL:LazyPIM:2017}
Boroumand, A., et~al.: {LazyPIM: An Efficient Cache Coherence Mechanism for
  Processing-in-Memory}.
\newblock IEEE Comput. Archit. Lett.  (2017)

\bibitem{Bress:GPUcoproc:IS:2013}
Bress, S., et~al.: Efficient co-processor utilization in database query
  processing.
\newblock Inf. Syst.  (2013)

\bibitem{Burger1996}
Burger, D., et~al.: {Memory bandwidth limitations of future microprocessors}.
\newblock In: Proc. ISCA (1996)

\bibitem{Chang:PhD:2017}
Chang, K.: Architectural techniques for improving nand flash memory
  reliability. doctoral dissertation. cmu (2017)

\bibitem{Cho2013}
Cho, S., et~al.: {Active disk meets flash}.
\newblock In: Proc. ICS (2013)

\bibitem{Choi2012}
Choi, Y., et~al.: {A 20nm 1.8V 8Gb PRAM with 40MB/s program bandwidth}.
\newblock In: 2012 IEEE Int. Solid-State Circuits Conf. (2012)

\bibitem{Coburn2013}
Coburn, J., Bunker, T., Schwarz, M., Gupta, R., Swanson, S.: {From ARIES to
  MARS}.
\newblock In: Proc. SOSP (2013)

\bibitem{ThinkingMachines}
Corporation, T.M.: Paris reference manual  (1991)

\bibitem{De2013}
De, A., et~al.: {Minerva: Accelerating Data Analysis in Next-Generation SSDs}.
\newblock In: 2013 IEEE 21st Annu. Int. Symp. Field-Programmable Cust. Comput.
  Mach. (2013)

\bibitem{Dennard1999}
Dennard, R.H., et~al.: {Design of Ion-Implanted MOSFETs with Very Small
  Physical Dimensions}  (1999)

\bibitem{DeWitt:CACM:DatabaseMachines:1992}
DeWitt, D., Gray, J.: Parallel database systems: The future of high performance
  database systems  (1992)

\bibitem{Draper2002}
Draper, J., et~al.: {The architecture of the DIVA processing-in-memory chip}.
\newblock In: Proc. ICS '02 (2002)

\bibitem{Drumond2017}
Drumond, M., et~al.: {The Mondrian Data Engine}.
\newblock ACM SIGARCH Comput. Archit. News  (2017)

\bibitem{Elliott1999}
Elliott, D., et~al.: {Computational RAM: implementing processors in memory}.
\newblock IEEE Des. Test Comput.  (1999)

\bibitem{Elliott2008}
Elliott, D., et~al.: {Computational Ram: A Memory-simd Hybrid And Its
  Application To Dsp}.
\newblock In: Proc. IEEE Cust. Integr. Circuits Conf. (2008)

\bibitem{Faggin1996}
Faggin, F., et~al.: {The history of the 4004}  (1996)

\bibitem{Farmahini-Farahani2015}
Farmahini-Farahani, A., et~al.: {NDA: Near-DRAM acceleration architecture
  leveraging commodity DRAM devices and standard memory modules}.
\newblock HPCA  (2015)

\bibitem{Fitch2009}
Fitch, B.G., et~al.: {Using the Active Storage Fabrics model to address
  petascale storage challenges}.
\newblock In: Proc. PDSW '09. New York, New York, USA (2009)

\bibitem{Francisco2011}
Francisco, P.: {The Netezza data appliance architecture: A platform for high
  performance data warehousing and analytics}.
\newblock IBM Redbooks  (2011)

\bibitem{Gao2015}
Gao, M., Ayers, G., Kozyrakis, C.: {Practical Near-Data Processing for
  In-Memory Analytics Frameworks}.
\newblock Proc. PACT  (2015)

\bibitem{Gao2016}
Gao, M., Kozyrakis, C.: {HRL: Efficient and flexible reconfigurable logic for
  near-data processing}.
\newblock Proc. - Int. Symp. High-Performance Comput. Archit.  (2016)

\bibitem{Gao2017}
Gao, M., et~al.: {TETRIS: Scalable and Efficient Neural Network Acceleration
  with 3D Memory}.
\newblock Asplos  (2017)

\bibitem{Mutlu2018}
Ghose, S., et~al.: {Enabling the Adoption of Processing-in-Memory: Challenges,
  Mechanisms, Future Research Directions}.
\newblock J. Phys. Chem. B  (2018)

\bibitem{Gokhale2015}
Gokhale, M., Lloyd, S., Hajas, C.: {Near memory data structure rearrangement}.
\newblock In: Proc. MEMSYS (2015)

\bibitem{Gokhale1995}
Gokhale, M., et~al.: {Processing in memory: the Terasys massively parallel PIM
  array}  (1995)

\bibitem{Gray:RulesOfThumb:ICDE:2000}
Gray, J., Shenoy, P.J.: Rules of thumb in data engineering.
\newblock In: Proc. ICDE (2000)

\bibitem{Gu2016}
Gu, B., et~al.: {Biscuit: A Framework for Near-Data Processing of Big Data
  Workloads}.
\newblock In: Proc. ISCA (2016)

\bibitem{Hadidi2017}
Hadidi, R., Nai, L., Kim, H., Kim, H.: {CAIRO}.
\newblock ACM Trans. Archit. Code Optim. \textbf{14}(4), 1--25 (2017)

\bibitem{Hall1999}
Hall, M., et~al.: {Mapping irregular applications to DIVA, a PIM-based
  data-intensive architecture}.
\newblock In: ACM/IEEE SC 1999 Conf. SC 1999 (1999)

\bibitem{Nikos:CS:TowardDarkSiliconInServers:2011}
Hardavellas, N., et~al.: Toward dark silicon in servers.
\newblock IEEE Micro  (2011)

\bibitem{Hong:ALT:PACT:2016}
Hong, B., et~al.: Accelerating linked-list traversal through near-data
  processing.
\newblock In: Proc. PACT (2016)

\bibitem{Hsieh2016}
Hsieh, K., et~al.: {Accelerating pointer chasing in 3D-stacked memory:
  Challenges, mechanisms, evaluation}.
\newblock Proc. ICCD  (2016)

\bibitem{Hsieh2016ISCA}
Hsieh, K., et~al.: {Transparent Offloading and Mapping (TOM): Enabling
  Programmer-Transparent Near-Data Processing in GPU Systems}.
\newblock Proc. ISCA  (2016)

\bibitem{Ist2017}
Istv{\'{a}}n, Z., et~al.: {Caribou}.
\newblock Proc. VLDB Endow.  (2017)

\bibitem{HBM}
JEDEC: High bandwidth memory {(HBM) DRAM}. {Standard No. JESD235B} (2018)

\bibitem{Kang2017}
Kang, D., et~al.: {256 Gb 3 b/Cell V-nand Flash Memory With 48 Stacked WL
  Layers}.
\newblock IEEE J. Solid-State Circuits  (2017)

\bibitem{YiKang1999}
Kang, Y., et~al.: {FlexRAM: toward an advanced intelligent memory system}.
\newblock In: Proc. VLSI (1999)

\bibitem{Kang2013}
Kang, Y., et~al.: {Enabling cost-effective data processing with smart SSD}.
\newblock In: 2013 IEEE 29th Symp. Mass Storage Syst. Technol., pp. 1--12. IEEE
  (2013)

\bibitem{Kaplan:JSFI:PIM:2017}
Kaplan, R., et~al.: From processing-in-memory to processing-in-storage.
\newblock Supercomputing Frontiers and Innovations  (2017)

\bibitem{Kaxiras:MLD:VectorIRAM:1997}
Kaxiras, S., et~al.: Distributed vector architecture: Beyond a single
  vector-iram.
\newblock In: In First Workshop on Mixing Logic and DRAM: Chips that Compute
  and Remember (1997)

\bibitem{Keeton:SigmodRec:IDISK:1998}
Keeton, K., et~al.: A case for intelligent disks (idisks).
\newblock SIGMOD Rec.  (1998)

\bibitem{Samsung1GB}
Kim, C., Cho, J., et~al.: 11.4 a 512gb 3b/cell 64-stacked wl 3d v-nand flash
  memory.
\newblock In: Proc. ISSCC (2017)

\bibitem{Kim2009}
Kim, D.H., et~al.: {TSV-aware interconnect length and power prediction for 3D
  stacked ICs}.
\newblock In: Proc. IIC (2009)

\bibitem{Kim2012}
Kim, J.S., et~al.: {A 1.2 V 12.8 GB/s 2 Gb Mobile Wide-I/O DRAM With 4 x 128
  I/Os Using TSV Based Stacking}.
\newblock IEEE J. Solid-State Circuits  (2012)

\bibitem{Moon:InfSci:ISC:2016}
Kim, S., et~al.: In-storage processing of database scans and joins.
\newblock Inf. Sci.  (2016)

\bibitem{Kogge1994}
Kogge, P.: {EXECUBE-A New Architecture for Scaleable MPPs}.
\newblock In: 1994 Int. Conf. Parallel Process. (1994)

\bibitem{Koo2017}
Koo, G., et~al.: {Summarizer}.
\newblock In: Proc. MICRO-50 '17 (2017)

\bibitem{Kozyrakis1997}
Kozyrakis, C., et~al.: {Scalable processors in the billion-transistor era:
  IRAM}.
\newblock Computer (Long. Beach. Calif).  (1997)

\bibitem{Lee2009}
Lee, B.C., et~al.: {Architecting phase change memory as a scalable dram
  alternative}.
\newblock In: Proc. ISCA (2009)

\bibitem{Lee2014}
Lee, D.U., et~al.: {25.2 A 1.2V 8Gb 8-channel 128GB/s high-bandwidth memory
  (HBM) stacked DRAM with effective microbump I/O test methods using 29nm
  process and TSV}.
\newblock In: Proc. ISSCC (2014)

\bibitem{Li:BitWeaving:SIGMOD:2013}
Li, Y., Patel, J.M.: Bitweaving: Fast scans for main memory data processing.
\newblock In: Proc. SIGMOD (2013)

\bibitem{Liu2018}
Liu, J., Zhao, H., Ogleari, M.A., Li, D., Zhao, J.: {Processing-in-memory for
  energy-efficient neural network training: A heterogeneous approach}.
\newblock Proc. MICRO  (2018)

\bibitem{Loh2013}
Loh, G., et~al.: {A Processing-in-Memory Taxonomy and a Case for Studying
  Fixed-function PIM}.
\newblock Wondp  (2013)

\bibitem{Masuoka1985}
Masuoka, F., et~al.: {A 256K flash EEPROM using triple polysilicon technology}.
\newblock In: Proc. ISSCC (1985)

\bibitem{Masuoka1987}
Masuoka, F., et~al.: {New ultra high density EPROM and flash EEPROM with NAND
  structure cell}.
\newblock In: 1987 Int. Electron Devices Meet. (1987)

\bibitem{Miller2011}
Miller, M.J.: {Bandwidth engine{\textregistered} serial memory chip breaks 2
  billion accesses/sec}.
\newblock In: Hot Chips (2011)

\bibitem{Ming2015}
Ming, S.w.J., et~al.: {BlueDBM: An Appliance for Big Data Analytics}.
\newblock Proc. ISCA  (2015)

\bibitem{Minutoli:RadixSortEmu1:WONDP:2015}
Minutoli, M., et~al.: Implementing radix sort on emu 1.
\newblock In: Proc. WoNDP (2015)

\bibitem{Minutoli2015}
Minutoli, M., et~al.: {Implementing Radix Sort on Emu 1}.
\newblock Work. Near-Data Process.  (2015)

\bibitem{Mirzadeh2015}
Mirzadeh, N.S., Kocberber, O., Falsafi, B., Ecocloud, B.G., Grot, B.: {Sort vs.
  Hash Join Revisited for Near-Memory Execution}.
\newblock 5th Work. Archit. Syst. Big Data (EPFL-CONF-209121) (2015)

\bibitem{Muramatsu:JCDL:IfYouBuildItWillTheyCome:2004}
Muramatsu, B., et~al.: If you build it, will they come?
\newblock Proc. JCDL  (2004)

\bibitem{Nai2017}
Nai, L., Hadidi, R., Sim, J., Kim, H., Kumar, P., Kim, H.: {GraphPIM: Enabling
  Instruction-Level PIM Offloading in Graph Computing Frameworks}.
\newblock Proc. HPCA  (2017)

\bibitem{Nair2015}
Nair, R., et~al.: {Active Memory Cube: A processing-in-memory architecture for
  exascale systems}  (2015)

\bibitem{Oskin1998}
Oskin, M., et~al.: {Active Pages: A Computation Model for Intelligent Memory}.
\newblock ACM SIGARCH  (1998)

\bibitem{Parat2015}
Parat, K., Dennison, C.: {A floating gate based 3D NAND technology with CMOS
  under array}.
\newblock In: Proc. IEDM (2015)

\bibitem{Park2015}
Park, K.T., et~al.: {Three-Dimensional 128 Gb MLC Vertical nand Flash Memory
  With 24-WL Stacked Layers and 50 MB/s High-Speed Programming}.
\newblock IEEE J. Solid-State Circuits  (2015)

\bibitem{Patterson:IRAM:Micro:1997}
Patterson, D., et~al.: A case for intelligent ram.
\newblock Micro  (1997)

\bibitem{Pawlowski2011}
Pawlowski, J.T.: {Hybrid memory cube (HMC)}.
\newblock In: 2011 IEEE Hot Chips 23 Symp. (2011)

\bibitem{Pugsley2014}
Pugsley, S.H., et~al.: {NDC: Analyzing the impact of 3D-stacked memory+logic
  devices on MapReduce workloads}.
\newblock In: Proc. ISPASS (2014)

\bibitem{Riedel1999}
Riedel, E., Nagle, D.: {Active Disks - Remote Execution for Network-Attached
  Storage Thesis Committee :}.
\newblock Science  (1999)

\bibitem{Riedel:VLDB:ActiveStorage:1998}
Riedel, E., et~al.: Active storage for large-scale data mining and multimedia.
\newblock In: Proc. VLDB (1998)

\bibitem{Riedel2001}
Riedel, E., et~al.: {Active disks for large-scale data processing}.
\newblock Computer (Long. Beach. Calif).  (2001)

\bibitem{Sakuma2013}
Sakuma, K., et~al.: {Highly Scalable Horizontal Channel 3-D NAND Memory
  Excellent in Compatibility With Conventional Fabrication Technology}.
\newblock IEEE Electron Device Lett.  (2013)

\bibitem{Schaller1997}
Schaller, R.: {Moore's law: past, present and future}.
\newblock IEEE Spectr.  (1997)

\bibitem{Scheuerlein2013}
Scheuerlein, R., et~al.: {A 130.7mm2 2-Layer 32Gb ReRAM Memory Device in 24nm
  Technology}.
\newblock Proc. IEEE Int. Solid-State Circuits Conf. Dig. Tech. Pap.  (2013)

\bibitem{Scrbak2017}
Scrbak, M., et~al.: {Exploring the Processing-in-Memory design space}.
\newblock J. Syst. Archit.  (2017)

\bibitem{Seshadri2014}
Seshadri, S., et~al.: {Willow: A User-Programmable SSD}.
\newblock Usenix, Osdi  (2014)

\bibitem{Seshadri2017}
Seshadri, V., Mowry, T.C., Lee, D., Mullins, T., Hassan, H., Boroumand, A.,
  Kim, J., Kozuch, M.A., Mutlu, O., Gibbons, P.B.: {Ambit}  (2017)

\bibitem{Seshadri2015}
Seshadri, V., et~al.: {Fast Bulk Bitwise AND and OR in DRAM}.
\newblock IEEE Comput. Archit. Lett.  (2015)

\bibitem{Shalf2015}
Shalf, J.M., Leland, R.: {Computing beyond Moore's Law}.
\newblock Computer (Long. Beach. Calif).  (2015)

\bibitem{Siegl:MEMSYS:PIMsurvey:2016}
Siegl, P., et~al.: Data-centric computing frontiers: A survey on
  processing-in-memory.
\newblock In: Proceedings MEMSYS (2016)

\bibitem{Silvagni2017}
Silvagni, A.: {3D NAND Flash Based on Planar Cells}.
\newblock Computers  (2017)

\bibitem{Strukov2008}
Strukov, D.B., et~al.: {The missing memristor found}.
\newblock Nature  (2008)

\bibitem{Swanson}
Swanson, S.: {Near Data Computation : It ' s Not ( Just ) About Performance}
  (2015)

\bibitem{Gray:SienceExponentialWorld:Nature:2006}
Szalay, A., Gray, J.: 2020 computing: Science in an exponential world  (2006)

\bibitem{Tang2017}
Tang, X., Kislal, O., Kandemir, M., Karakoy, M.: {Data movement aware
  computation partitioning}  (2017)

\bibitem{Tiwari:ActiveDiskInSitu:FAST:2013}
Tiwari, D., et~al.: Active flash: Towards energy-efficient, in-situ data
  analytics on extreme-scale machines.
\newblock In: Proc. FAST (2013)

\bibitem{Torrellas:FlexRAMretro:ICCD:2012}
Torrellas, J.: Flexram: Toward an advanced intelligent memory system: A
  retrospective paper.
\newblock In: Proc. ICCD (2012)

\bibitem{Tsai2018}
Tsai, P.A., Chen, C., Sanchez, D.: {Adaptive scheduling for systems with
  asymmetric memory hierarchies}.
\newblock Proc. MICRO  (2018)

\bibitem{Vermij2017}
Vermij, E., et~al.: {Sorting big data on heterogeneous near-data processing
  systems}.
\newblock In: Proc. CF (2017)

\bibitem{Vijaykumar2018}
Vijaykumar, N., et~al.: {A Case for Richer Cross-Layer Abstractions: Bridging
  the Semantic Gap with Expressive Memory}.
\newblock In: Proc. ISCA (2018)

\bibitem{Villa2010}
Villa, C., et~al.: {A 45nm 1Gb 1.8V phase-change memory}.
\newblock In: Proc. ISSCC (2010)

\bibitem{Wang:SSDLists:DAMON:2016}
Wang, J., Park, D., Kee, Y.S., Papakonstantinou, Y., Swanson, S.: Ssd
  in-storage computing for list intersection.
\newblock In: Proc. DaMoN (2016)

\bibitem{Wang2016}
Wang, J., et~al.: {SSD in-storage computing for list intersection}.
\newblock In: Proc. DaMoN (2016)

\bibitem{Wang2015}
Wang, Y., et~al.: {ProPRAM: Exploiting the transparent logic resources in
  Non-Volatile Memory for Near Data Computing}.
\newblock Proc. DAC  (2015)

\bibitem{Willhalm:SIMDScan:VLDB:2009}
Willhalm, T., et~al.: Simd-scan: Ultra fast in-memory table scan using on-chip
  vector processing units.
\newblock Proc. VLDB Endow.  (2009)

\bibitem{Wong2012}
Wong, H.S.P., et~al.: {Metalâ€“Oxide RRAM}.
\newblock Proc. IEEE  (2012)

\bibitem{Teubner:IBEX:SIGMOD:2013}
Woods, L., et~al.: Less watts, more performance: An intelligent storage engine
  for data appliances.
\newblock In: Proc. SIGMOD (2013)

\bibitem{Teubner:IBEX:VLDB:2014}
Woods, L., et~al.: Ibex: An intelligent storage engine with support for
  advanced sql offloading.
\newblock Proc. VLDB  (2014)

\bibitem{Wu2014}
Wu, L., et~al.: {Q100}.
\newblock In: Proc. ASPLOS (2014)

\bibitem{WulfWmAandMcKee1994}
Wulf, W.A., McKee, S.A.: {Hitting the Memory Wall : Implications of the
  Obvious}.
\newblock SIGARCH  (1994)

\bibitem{Xi2015}
Xi, S.L., et~al.: {Beyond the Wall: Near-Data Processing for Databases}.
\newblock Proc. DaMoN  (2015)

\bibitem{Xie2017}
Xie, C., Song, S.L., Wang, J., Zhang, W., Fu, X.: {Processing-in-Memory Enabled
  Graphics Processors for 3D Rendering}.
\newblock Proc. HPCA  (2017)

\bibitem{Zhang2014}
Zhang, D., et~al.: {Top-Pim}.
\newblock Proc. HPDC '14  (2014)

\bibitem{Zhang:NPP:MSPC:2013}
Zhang, D.P., et~al.: A new perspective on processing-in-memory architecture
  design.
\newblock In: Proc. MSPS (2013)

\bibitem{Zhang2018}
Zhang, M., Zhuo, Y., Wang, C., Gao, M., Wu, Y., Chen, K., Kozyrakis, C., Qian,
  X.: {GraphP: Reducing Communication for PIM-Based Graph Processing with
  Efficient Data Partition}.
\newblock Proc. HPCA \textbf{2018-Febru} (2018)

\bibitem{Huai2008}
Zhao, W., et~al.: {Spin transfer torque (STT)-MRAM--based runtime
  reconfiguration FPGA circuit}.
\newblock Proc. TECS  (2009)

\bibitem{Zhu2013}
Zhu, Q., et~al.: {A 3D-stacked logic-in-memory accelerator for
  application-specific data intensive computing}.
\newblock In: 2013 IEEE Int. 3D Syst. Integr. Conf. (2013)

\bibitem{Zhu2013HPEC}
Zhu, Q., et~al.: {Accelerating sparse matrix-matrix multiplication with
  3D-stacked logic-in-memory hardware}.
\newblock In: Proc. HPECC (2013)

\bibitem{Teich:FPGAsql:TRTS:2016}
Ziener, D., et~al.: Fpga-based dynamically reconfigurable sql query processing.
\newblock ACM TRTS  (2016)

\end{thebibliography}

\end{document}